\documentclass{article}
\usepackage[hidelinks, colorlinks=true]{hyperref}
\usepackage[english]{babel}
\usepackage{amsmath,amssymb}
\usepackage[letterpaper,top=2cm,bottom=2cm,left=3cm,right=3cm,marginparwidth=1.75cm]{geometry}
\usepackage{authblk}
\usepackage{dsfont}
\newcommand{\id}{\mathds{1}}

\newcommand{\ket}[1]{\vert #1 \rangle}
\newcommand{\bra}[1]{\langle #1 \vert}

\usepackage[dvipsnames]{xcolor}
\usepackage{tikz}
\usetikzlibrary{decorations.pathmorphing,decorations.markings,arrows,arrows.meta,shapes,patterns.meta}

\tikzset{
  every path/.style={line cap=round},
  bevel/.style={preaction={draw,white,line width=2pt,line cap=round}},
  thick bevel/.style={preaction={draw,white,line width=4pt,line cap=round}},
  disk/.style={circle,draw=black,fill=black},
  fdisk/.style={circle,draw=black},
  square/.style={draw=black,fill=black},
  fsquare/.style={draw=black},
  star8/.style={draw=black,star,star points=8,fill=black},
  fstar8/.style={draw=black,star,star points=8},
  disk normal/.style={disk,inner sep=1.45pt},
  disk large/.style={disk,inner sep=1.75pt},
  disk small/.style={disk,inner sep=1pt},
  disk tiny/.style={disk,inner sep=0.7pt},
  fdisk normal/.style={fdisk,inner sep=1.45pt},
  fdisk large/.style={fdisk,inner sep=1.75pt},
  fdisk small/.style={fdisk,inner sep=1pt},
  fdisk tiny/.style={fdisk,inner sep=1.45pt},
  star8 normal/.style={star8,star point height=0.25mm,inner sep=1.1pt},
  star8 small/.style={star8,star point height=0.2mm,inner sep=0.9pt},
  star8 large/.style={star8,star point height=0.25mm,inner sep=1.45pt},
  star8 tiny/.style={star8,star point height=0.2mm,inner sep=0.7pt},
  fstar8 normal/.style={fstar8,star point height=0.25mm,inner sep=1.1pt},
  fstar8 small/.style={fstar8,star point height=0.2mm,inner sep=0.9pt},
  fstar8 large/.style={fstar8,star point height=0.25mm,inner sep=1.45pt},
  fstar8 tiny/.style={fstar8,star point height=0.2mm,inner sep=0.7pt},
  square normal/.style={square, inner sep=1.1pt},
  square small/.style={square, inner sep=0.9pt},
  square large/.style={square, inner sep=1.45pt},
  square tiny/.style={square, inner sep=0.7pt},
  fsquare normal/.style={fsquare, inner sep=1.1pt},
  fsquare small/.style={fsquare, inner sep=0.9pt},
  fsquare large/.style={fsquare, inner sep=1.75pt},
  fsquare tiny/.style={fsquare, inner sep=0.7pt},
  -mid/.style={
    decoration={markings,mark=at position 0.5*\pgfdecoratedpathlength+0.6*3pt with \arrow{>[width=3pt]}},
    postaction={decorate}
  },
  mid-/.style={
    decoration={markings,mark=at position 0.5*\pgfdecoratedpathlength+0.6*3pt with \arrow{<[width=3pt]}},
    postaction={decorate}
  }
}
\tikzdeclarepattern{name=rising,bounding box={(0,-0.75pt) and (2pt,0.75pt)},tile size={(2pt,2pt)},tile transformation={rotate=45},code={\draw [line width=1.5pt] (0,0) -- (2pt,0);}}
\tikzdeclarepattern{name=falling,bounding box={(0,-0.75pt) and (2pt,0.75pt)},tile size={(2pt,2pt)},tile transformation={rotate=-45},code={\draw [line width=1.5pt] (0,0) -- (2pt,0);}}

\title{On two-dimensional tensor network group symmetries}

\author[1]{Jos\'e Garre-Rubio}
\author[2]{Andr\'as Moln\'ar}

\affil[1]{\small Instituto de F\'isica Te\'orica, UAM/CSIC, C. Nicol\'as Cabrera 13-15, Cantoblanco, 28049 Madrid, Spain}
\affil[2]{\small University of Vienna, Faculty of Mathematics, Oskar-Morgenstern-Platz 1, 1090 Vienna, Austria}

\date{}
\begin{document}

\maketitle

\begin{abstract}
We introduce two-dimensional tensor-network representations of finite groups carrying a 4‑cocycle index. We characterize the associated gapped (2+1)D phases that emerge when these anomalous symmetries act on tensor‑network ground states. We further develop related tensor‑network unitaries that generate symmetric states representing (3+1)D symmetry-protected topological phases. Although aspects of these constructions have been previously addressed, our contribution unifies them within a single tensor‑network framework and emphasizes the explicit formulation of local tensor equations encoding global consistency conditions.
\end{abstract}

\section{Introduction}

Symmetries—and particularly their anomalies—are central to the classification of quantum phases in many-body systems. Tensor networks (TN) \cite{reviewPEPS} offer a powerful toolkit to encode even non-group symmetries \cite{Lootens21A,Molnar22,Delcamp_2022} and to represent their symmetric ground states. The essential idea is that global properties of a state can be captured locally via patterns in the tensors and their virtual indices.

A prominent application of TN is in the description and classification of symmetry-protected topological (SPT) phases. By definition, these phases cannot be continuously connected to a trivial phase through gapped, symmetry-preserving Hamiltonians \cite{Hastings05}, and hence lie outside Landau's symmetry-breaking paradigm. The classification of 1D SPT phases was unlocked through matrix-product states (MPS) \cite{Fannes92,PerezGarcia07} and the understanding of their characteristic degenerate entanglement spectra \cite{Pollmann10}. Subsequent works provided a complete classification in (1+1)D using group cohomology methods \cite{Chen11,Schuch11,ogata2021classification}.  

This algebraic structure—captured by group cohomology—extends to higher dimensions, where (d+1)D SPT phases are classified by $H^{d+1}(G,U(1))$, manifesting as anomalies of d-dimensional symmetries \cite{Chen13}. In this context, a symmetry is called \emph{anomalous} if it cannot be gauged, equivalently, if there is no symmetric, short-range entangled state invariant under it \cite{kapustin14}.

In an earlier work \cite{Chen11A} from the same authors of Ref.~\cite{Chen13}, the 1D anomaly operator—classified by the third cohomology group of $G$ $H^3(G,U(1))$—was explicitly realized as a matrix-product operator, which then became the local on-site symmetry operator of a 2D SPT model. Our goal is to generalize this construction to higher dimensions: we aim to encode 2D anomalous symmetries labeled by $H^4(G,U(1))$, in order to study (3+1)D SPT phases. Concretely, our contributions are three-fold:
\begin{itemize}
  \item We develop a tensor-network framework to embed a 4‑cocycle into a 2D representation of a finite group. We derive sufficient local tensor conditions ensuring that the corresponding global group relations hold.
  
  \item We employ this framework to classify phases invariant under these anomalous symmetries. By analyzing how the TN operator acts on a TN state, we derive equations whose distinct solutions correspond to different phases—a strategy analogous to the one used in \cite{Garre22_MPOSYM} for one-dimensional systems. This allows us to show that no anomalous symmetry, with a nontrivial 4‑cocycle, admits a unique, symmetric ground state.
  \item Finally, we translate our TN operators into unitaries representing 2D group symmetries, which serve as on-site operators for a 3D global symmetry of SPT models. From this TN perspective, we can see that these on-site operators fuse at the virtual level and propagate the anomaly to the 2D boundary.
\end{itemize}

We note that in Ref.~\cite{Delcamp_2022} the TN representation with trivial $4$-cocycle is studied. The concept of symmetry based on groups, as such as of SPT order, has been generalized to categorical symmetries \cite{Gaiotto_2015} and the classification of its symmetric gapped phases has been achieved \cite{Thorngren19,Garre22_MPOSYM}. More recently, several works have extended these ideas to two-dimensional systems \cite{inamura202521dlatticemodelstensor,bhardwaj2024gappedphases21dnoninvertible,bhardwaj2025gappedphases21dnoninvertible,Delcamp_2024,Inamura_2024} where the symmetry is given by a fusion 2-category. In categorical terms we analyze the $\mathsf{2Vec}_G^\omega$ case and its modules 2-categories.

{\it Noted Added:} As this paper was nearing completion, Ref.~\cite{kawagoe25} appeared, proposing a method to compute the 4‑cocycle index for 2D unitary quantum-circuit symmetries.

\section{Tensor network representation of finite groups in 2D}\label{TNOG}

Let us consider tensor network operators defined on the vertices of a two-dimensional hexagonal lattice. These operators are labeled by the elements of a finite group, $g\in G$, so that they are constructed with two tensors $T_g$ and $T'_g$, one for each type of triangular vertex:

\begin{equation}\label{Og}
O_g = 
\begin{tikzpicture}[scale=0.75, baseline=2cm]
    \draw[RoyalBlue!80, very thick] (5.225, 2.65) -- (5.495, 1.918);
    \draw[RoyalBlue!80, very thick] (4.381, 2.073) -- (4.65, 1.341);
    \draw[RoyalBlue!80, very thick] (4.366, 4.336) -- (4.635, 3.604);
    \draw[RoyalBlue!80, very thick] (5.214, 3.787) -- (5.483, 3.055);
    \draw[RoyalBlue!80, very thick] (3.52, 2.652) -- (3.79, 1.92);
    \draw[RoyalBlue!80, very thick] (3.544, 3.747) -- (3.813, 3.015);
    \draw[very thick] (3.669, 3.387) -- (3.669, 2.258) -- (4.516, 1.693) -- (5.362, 2.258) -- (5.362, 3.387) -- (4.516, 3.951) -- cycle;
    \draw[very thick] (4.516, 3.951) -- (4.516, 4.516);
    \draw[very thick] (3.669, 3.387) -- (2.822, 3.951);
    \draw[very thick] (5.362, 3.387) -- (6.209, 3.951);
    \draw[very thick] (5.362, 2.258) -- (6.209, 1.693);
    \draw[very thick] (3.669, 2.258) -- (2.822, 1.693);
    \draw[very thick] (4.516, 1.693) -- (4.516, 1.129);
    \node[disk normal] at (3.669, 3.387) {};
    \node[disk normal] at (4.516, 3.951) {};
    \node[disk normal] at (5.362, 3.387) {};
    \node[disk normal] at (3.669, 2.258) {};
    \node[disk normal] at (4.516, 1.693) {};
    \node[disk normal] at (5.362, 2.258) {};
    \node[anchor=center] at (4.1, 3.2) {$T_g$};
    \node[anchor=center] at (4.9, 4) {$T'_g$};
\end{tikzpicture} \ ;
\end{equation}
for simplicity, we think of the operators $O_g$ as if they were defined on an infinite plane. 

{\bf Product.} These operators form a representation of the group $G$, i.e., $O_g O_h = O_{gh}$ for all $g,h\in G$, if the following two conditions are satisfied:
\begin{enumerate}
    \item The product of the tensors $T_g$ and $T_h$ is related to the tensor $T_{gh}$ by applying a {\it fusion matrix product operator} on the virtual indices of the tensors. This fusion MPO is constructed from a five-index tensor, denoted as $F_{g,h}$, such that
\begin{equation}\label{prodTgh}
\begin{tikzpicture}[scale=1, baseline=3cm]
    \draw[very thick] (-5.422, 4.927) -- (-4.293, 4.08) -- (-3.164, 4.927) -- (-3.164, 4.927);
    \draw[very thick] (-4.293, 4.08) -- (-4.857, 2.951) -- (-4.857, 2.951);
    \draw[RoyalBlue!80, very thick] (-4.293, 4.645) -- (-4.293, 3.516) -- (-4.293, 3.233);
    \node[disk normal] at (-4.293, 4.08) {};
    \draw[very thick] (-5.422, 3.516) -- (-4.293, 2.669) -- (-3.164, 3.516) -- (-3.164, 3.516);
    \draw[very thick] (-4.293, 2.669) -- (-4.857, 1.54) -- (-4.857, 1.54);
    \draw[RoyalBlue!80, very thick] (-4.293, 3.233) -- (-4.293, 2.105) -- (-4.293, 1.822);
    \node[disk normal] at (-4.293, 2.669) {};
    \draw[shift={(-0.428, 3.43)}, scale=1.25, very thick] (0, 0) -- (-0.564, -1.129) -- (-0.564, -1.129);
    \draw[RoyalBlue!80, very thick] (-0.428, 3.995) -- (-0.428, 2.866) -- (-0.428, 2.583);
    \node[disk normal] at (-0.428, 3.43) {};
    \node[anchor=center, font=\LARGE] at (-2.583, 3.227) {$=$};
    \draw[shift={(0.924, 4.134)}, xscale=1.25, RedOrange, very thick] (0, 0) .. controls (0.047, -0.706) and (-1.082, -2.117) .. (-1.693, -2.117) .. controls (-2.305, -2.117) and (-2.399, -0.706) .. (-1.834, 0) .. controls (-1.27, 0.706) and (-0.047, 0.706) .. cycle;
    \draw[very thick] (-0.428, 3.43) -- (-1.557, 3.995);
    \draw[very thick] (-0.428, 3.43) -- (0.924, 4.136) -- (0.924, 4.136);
    \node[disk normal] at (-1.51, 3.971) {};
    \node[disk normal] at (-1.133, 2.022) {};
    \node[disk normal] at (0.924, 4.136) {};
    \draw[very thick] (1.265, 4.559) -- (0.924, 4.136) -- (1.265, 3.712) -- (1.265, 3.712);
    \draw[very thick] (-1.83, 4.579) -- (-1.498, 3.98) -- (-1.96, 3.664);
    \draw[very thick] (-1.937, 1.928) -- (-1.138, 2.06) -- (-1.3, 1.436);
    \node[anchor=center] at (-3.937, 3.805) {$T_g
$};
    \node[anchor=center] at (-3.937, 2.394) {$T_h$};
    \node[anchor=center] at (-0.035, 3.212) {$T_{gh}$};
    \node[anchor=center] at (-1.263, 3.545) {$F_{g,h}$};
    \node[anchor=center] at (-0.184, 4.924) {$g,h
$};
\end{tikzpicture}\ .
\end{equation}
The virtual level of this MPO is just dubbed as $g,h$ and it is depicted in red. A similar equation is satisfied for the tensors $T'_g$ placed at the vertices where the triangle is pointing upside down.

\item The tensors of the fusion MPO satisfy the orthogonality relation: $F_{g,h}F_{k,l}= \delta_{g,k}\delta_{h,l} \id_{g,h}\id_{gh} $, which can be depicted as follows:

\begin{equation}\label{orthoF}
\begin{tikzpicture}[scale=1, baseline=-2cm]
    \draw[RedOrange, very thick] (-4.165, -0.513) -- (-4.165, -3.053);
    \draw[RedOrange, very thick] (-2.949, -0.53) -- (-2.949, -3.07);
    \node[disk normal] at (-4.151, -1.775) {};
    \node[disk normal] at (-2.953, -1.787) {};
    \draw[very thick] (-4.137, -1.778) -- (-4.701, -1.778) -- (-4.983, -1.778);
    \draw[very thick] (-2.116, -1.777) -- (-2.681, -1.777) -- (-2.963, -1.777);
    \draw[very thick] (-4.131, -1.765) -- (-3.514, -1.142) -- (-2.946, -1.778) -- (-3.564, -2.402) -- cycle;
    \node[anchor=center, font=\LARGE] at (-1.725, -1.764) {$=$};
    \draw[shift={(1.686, -1.738)}, xscale=2.088, yscale=0.164, very thick] (0, 0) -- (-0.564, 0) -- (-0.847, 0);
    \draw[RedOrange, very thick] (0.234, -0.745) .. controls (0.221, -1.572) and (1.329, -1.59) .. (1.348, -0.748);
    \draw[shift={(1.396, -2.747)}, rotate=-179.913, RedOrange, very thick] (0, 0) .. controls (-0.013, -0.827) and (1.095, -0.845) .. (1.115, -0.003);
    \node[anchor=center] at (-4.569, -0.713) {$g,h$};
    \node[anchor=center] at (-2.539, -0.709) {$k,l$};
    \node[anchor=center] at (-0.86, -1.742) {$\delta_{g,k}\delta_{h,l}$};
    \node[anchor=center] at (0.831, -1.029) {$g,h$};
    \node[anchor=center] at (0.828, -2.49) {$g,h$};
    \node[anchor=center] at (-4.825, -1.585) {$gh$};
    \node[anchor=center] at (-2.544, -1.573) {$kl$};
    \node[anchor=center] at (1.925, -1.738) {$gh$};
    \node[anchor=center] at (-3.923, -1.299) {$g$};
    \node[anchor=center] at (-3.933, -2.274) {$h$};
    \node[anchor=center] at (-3.155, -1.27) {$k$};
    \node[anchor=center] at (-3.161, -2.274) {$l$};
\end{tikzpicture} \ .
\end{equation}
\end{enumerate}
Note that Eq.~\eqref{orthoF} allows us to concatenate Eq.~\eqref{prodTgh} in such a way that when multiplying a patch of tensors the fusion MPO is pushed to the boundary:

\begin{tikzpicture}[scale=1]
    \draw[shift={(6.103, 0.054)}, xscale=0.531, yscale=0.519, RoyalBlue!80, thick] (0, 0) -- (0.347, -1.068);
    \draw[shift={(3.846, 0.054)}, xscale=0.584, yscale=0.605, RoyalBlue!80, thick] (0, 0) -- (0.347, -1.068);
    \draw[shift={(2.717, 1.188)}, xscale=0.584, yscale=0.605, RoyalBlue!80, thick] (0, 0) -- (0.347, -1.068);
    \draw[shift={(7.232, 1.183)}, xscale=0.584, yscale=0.605, RoyalBlue!80, thick] (0, 0) -- (0.347, -1.068);
    \draw[shift={(7.232, 2.312)}, xscale=0.584, yscale=0.605, RoyalBlue!80, thick] (0, 0) -- (0.347, -1.068);
    \draw[shift={(3.846, 3.441)}, xscale=0.584, yscale=0.605, RoyalBlue!80, thick] (0, 0) -- (0.347, -1.068);
    \draw[shift={(4.975, 2.312)}, xscale=0.584, yscale=0.605, RoyalBlue!80, thick] (0, 0) -- (0.347, -1.068);
    \draw[shift={(4.975, 1.183)}, xscale=0.584, yscale=0.605, RoyalBlue!80, thick] (0, 0) -- (0.347, -1.068);
    \draw[shift={(2.733, 2.291)}, xscale=0.584, yscale=0.605, RoyalBlue!80, thick] (0, 0) -- (0.347, -1.068);
    \draw[shift={(8.572, 0.726)}, xscale=1.169, yscale=1.508, RedOrange, thick] (0, 0) .. controls (-0.698, -0.963) and (-4.386, -1.415) .. (-5.532, -0.679) .. controls (-6.679, 0.057) and (-5.284, 1.982) .. (-3.44, 2.208) .. controls (-1.596, 2.435) and (0.698, 0.963) .. cycle;
    \draw[RoyalBlue!80, thick] (-4.632, 0.15) -- (-4.285, -0.917);
    \draw[RoyalBlue!80, thick] (-3.484, 1.214) -- (-3.137, 0.146);
    \draw[RoyalBlue!80, thick] (-2.373, 0.096) -- (-2.027, -0.971);
    \draw[RoyalBlue!80, thick] (-1.223, 1.212) -- (-0.876, 0.144);
    \draw[RoyalBlue!80, thick] (-1.252, 2.382) -- (-0.905, 1.315);
    \draw[RoyalBlue!80, thick] (-2.347, 3.418) -- (-2, 2.351);
    \draw[RoyalBlue!80, thick] (-3.488, 2.369) -- (-3.141, 1.301);
    \draw[RoyalBlue!80, thick] (-4.621, 3.504) -- (-4.274, 2.436);
    \draw[RoyalBlue!80, thick] (-5.73, 1.169) -- (-5.383, 0.102);
    \draw[RoyalBlue!80, thick] (-5.776, 2.446) -- (-5.429, 1.378);
    \draw[thick] (-4.516, 3.104) -- (-5.644, 1.976) -- (-5.644, 0.847) -- (-4.516, -0.282) -- (-3.387, 0.847) -- (-3.387, 1.976) -- (-4.516, 3.104) -- (-4.516, 4.233);
    \draw[thick] (-3.387, 1.976) -- (-2.258, 3.104) -- (-2.258, 4.233) -- (-2.258, 4.233);
    \draw[thick] (-2.258, 3.104) -- (-1.129, 1.976) -- (-1.129, 0.847) -- (-2.258, -0.282) -- (-3.387, 0.847);
    \draw[thick] (-1.129, 1.976) -- (0, 3.104);
    \draw[thick] (-1.129, 0.847) -- (0, -0.282) -- (0, -0.282);
    \draw[thick] (-2.258, -0.282) -- (-2.258, -1.411) -- (-2.258, -1.411);
    \draw[thick] (-4.516, -0.282) -- (-4.516, -1.411);
    \draw[thick] (-5.644, 0.847) -- (-6.773, -0.282);
    \draw[thick] (-5.644, 1.976) -- (-6.773, 3.104);
    \node[disk normal] at (-5.644, 1.976) {};
    \node[disk normal] at (-5.644, 0.847) {};
    \node[disk normal] at (-4.516, 3.104) {};
    \node[disk normal] at (-3.387, 1.976) {};
    \node[disk normal] at (-4.516, -0.282) {};
    \node[disk normal] at (-3.387, 0.847) {};
    \node[disk normal] at (-2.258, 3.104) {};
    \node[disk normal] at (-1.129, 1.976) {};
    \node[disk normal] at (-2.258, -0.282) {};
    \node[disk normal] at (-1.129, 0.847) {};
    \draw[thick] (3.951, 3.104) -- (2.822, 1.976) -- (2.822, 0.847) -- (3.951, -0.282) -- (5.08, 0.847) -- (5.08, 1.976) -- (3.951, 3.104) -- (3.951, 4.233);
    \draw[thick] (5.08, 1.976) -- (6.209, 3.104) -- (6.209, 4.233) -- (6.209, 4.233);
    \draw[thick] (6.209, 3.104) -- (7.338, 1.976) -- (7.338, 0.847) -- (6.209, -0.282) -- (5.08, 0.847);
    \draw[shift={(7.338, 1.976)}, xscale=0.936, yscale=0.913, thick] (0, 0) -- (1.129, 1.129);
    \draw[shift={(7.338, 0.847)}, xscale=0.823, yscale=0.812, thick] (0, 0) -- (1.129, -1.129) -- (1.129, -1.129);
    \draw[shift={(6.209, -0.28)}, xscale=0.361, yscale=0.675, thick] (0, 0) -- (0, -1.129) -- (0, -1.129);
    \draw[shift={(3.951, -0.282)}, yscale=0.809, thick] (0, 0) -- (0, -1.129);
    \draw[thick] (2.822, 0.847) -- (1.693, -0.282);
    \draw[shift={(2.822, 1.976)}, xscale=0.71, yscale=0.717, thick] (0, 0) -- (-1.129, 1.129);
    \node[disk normal] at (2.822, 1.976) {};
    \node[disk normal] at (2.822, 0.847) {};
    \node[disk normal] at (3.951, 3.104) {};
    \node[disk normal] at (5.08, 1.976) {};
    \node[disk normal] at (3.951, -0.282) {};
    \node[disk normal] at (5.08, 0.847) {};
    \node[disk normal] at (6.209, 3.104) {};
    \node[disk normal] at (7.338, 1.976) {};
    \node[disk normal] at (6.209, -0.282) {};
    \node[disk normal] at (7.338, 0.847) {};
    \draw[thick] (-4.393, 2.826) -- (-5.522, 1.697) -- (-5.522, 0.568) -- (-4.393, -0.561) -- (-3.265, 0.568) -- (-3.265, 1.697) -- (-4.393, 2.826) -- (-4.393, 3.955);
    \draw[thick] (-3.265, 1.697) -- (-2.136, 2.826) -- (-2.136, 3.955) -- (-2.136, 3.955);
    \draw[thick] (-2.136, 2.826) -- (-1.007, 1.697) -- (-1.007, 0.568) -- (-2.136, -0.561) -- (-3.265, 0.568);
    \draw[thick] (-1.007, 1.697) -- (0.122, 2.826);
    \draw[thick] (-1.007, 0.568) -- (0.122, -0.561) -- (0.122, -0.561);
    \draw[thick] (-2.136, -0.561) -- (-2.136, -1.69) -- (-2.136, -1.69);
    \draw[thick] (-4.393, -0.561) -- (-4.393, -1.69);
    \draw[thick] (-5.522, 0.568) -- (-6.651, -0.561);
    \draw[thick] (-5.522, 1.697) -- (-6.651, 2.826);
    \node[disk normal] at (-5.522, 1.697) {};
    \node[disk normal] at (-5.522, 0.568) {};
    \node[disk normal] at (-4.393, 2.826) {};
    \node[disk normal] at (-3.265, 1.697) {};
    \node[disk normal] at (-4.393, -0.561) {};
    \node[disk normal] at (-3.265, 0.568) {};
    \node[disk normal] at (-2.136, 2.826) {};
    \node[disk normal] at (-1.007, 1.697) {};
    \node[disk normal] at (-2.136, -0.561) {};
    \node[disk normal] at (-1.007, 0.568) {};
    \node[anchor=center] at (-5.551, 2.27) {$g$};
    \node[anchor=center] at (-5.248, 1.607) {$h$};
    \node[disk normal] at (2.257, 2.541) {};
    \node[disk normal] at (1.896, -0.079) {};
    \node[disk normal] at (3.951, -0.842) {};
    \node[disk normal] at (6.209, -0.612) {};
    \node[disk normal] at (8.029, 0.156) {};
    \node[disk normal] at (8.062, 2.699) {};
    \node[disk normal] at (6.209, 3.883) {};
    \node[disk normal] at (3.951, 3.893) {};
    \draw[thick] (2.259, 2.544) -- (2.209, 2.834) -- cycle;
    \draw[thick] (3.951, 3.893) -- (4.096, 4.166) -- cycle;
    \draw[thick] (6.209, 3.883) -- (6.379, 4.093) -- cycle;
    \draw[thick] (1.896, -0.079) -- (1.878, -0.354) -- cycle;
    \draw[thick] (3.951, -0.842) -- (3.828, -1.047) -- cycle;
    \draw[thick] (6.209, -0.612) -- (6.095, -0.855) -- cycle;
    \draw[thick] (8.034, 0.159) -- (8.081, -0.075) -- cycle;
    \draw[thick] (8.07, 2.69) -- (8.357, 2.756) -- cycle;
    \draw[shift={(6.103, 3.441)}, xscale=0.584, yscale=0.605, RoyalBlue!80, thick] (0, 0) -- (0.347, -1.068);
    \node[anchor=center] at (3.15, 1.914) {$gh$};
    \node[anchor=center, font=\Large] at (0.804, 1.352) {$=$};
\end{tikzpicture}

{\bf Associativity.} We now consider the two equivalent ways of decomposing the product of three tensors:  $(T_g T_h) T_k = T_g( T_h T_k)$. We assume that the tensors $T_g$ for all $g\in G$ are left-invertible, i.e. there exists some tensor $T^{-1}_g$ acting on the physical space such that:

\begin{equation}
    \begin{tikzpicture}[scale=1, baseline=2cm]
    \draw[very thick] (-4.857, 4.08) -- (-3.728, 3.233) -- (-2.599, 4.08) -- (-2.599, 4.08);
    \draw[very thick] (-3.728, 3.233) -- (-4.293, 2.105) -- (-4.293, 2.105);
    \draw[very thick] (-4.857, 2.669) -- (-3.728, 1.822) -- (-2.599, 2.669) -- (-2.599, 2.669);
    \draw[very thick] (-3.728, 1.822) -- (-4.293, 0.693) -- (-4.293, 0.693);
    \node[anchor=center, font=\LARGE] at (-2.019, 2.38) {$=$};
    \node[anchor=center] at (-3.372, 2.958) {$T_g
$};
    \node[anchor=center] at (-3.252, 1.708) {$T^{-1}_g$};
    \draw[RoyalBlue!80, thick] (-3.728, 3.233) .. controls (-3.756, 4.137) and (-3.345, 4.13) .. (-3.101, 3.834) .. controls (-2.858, 3.539) and (-2.782, 2.954) .. (-2.807, 2.422) .. controls (-2.832, 1.889) and (-2.958, 1.409) .. (-3.116, 1.204) .. controls (-3.273, 0.999) and (-3.463, 1.069) .. (-3.577, 1.275) .. controls (-3.691, 1.481) and (-3.728, 1.822) .. (-3.728, 3.233);
    \node[disk normal] at (-3.728, 3.233) {};
    \node[disk normal] at (-3.728, 1.822) {};
    \draw[very thick] (0.304, 3.762) .. controls (-0.37, 2.934) and (0.437, 2.082) .. (1.014, 3.136);
    \draw[very thick] (-1.108, 3.771) .. controls (-0.447, 3.459) and (-0.52, 3.01) .. (-0.64, 2.779) .. controls (-0.76, 2.549) and (-0.927, 2.538) .. (-1.153, 2.86);
    \draw[very thick] (-0.986, 2.003) .. controls (-0.599, 2.618) and (-0.383, 2.54) .. (-0.269, 2.378) .. controls (-0.154, 2.217) and (-0.142, 1.971) .. (-0.164, 1.787) .. controls (-0.185, 1.604) and (-0.242, 1.482) .. (-0.62, 1.174);
\end{tikzpicture}
\end{equation}
Then, the associativity for any patch of tensors implies that the following concatenation of fusion MPOs for any size is equal:

\begin{equation}\label{assofmpo}
\begin{tikzpicture}[scale=1, baseline=2cm]
    \draw[shift={(-5.362, 5.08)}, yscale=1.333, RedOrange, very thick] (0, 0) -- (0, -3.387) -- (0, -3.387);
    \draw[shift={(-4.516, 5.08)}, yscale=1.333, RedOrange, very thick] (0, 0) -- (0, -3.387);
    \draw[very thick] (-3.951, 4.798) -- (-4.516, 4.233) -- (-3.951, 3.669);
    \draw[very thick] (-4.516, 4.233) -- (-5.362, 3.669) -- (-4.798, 3.104) -- (-4.798, 3.104);
    \draw[very thick] (-5.362, 3.669) -- (-5.927, 3.669) -- (-6.209, 3.669);
    \node[disk normal] at (-5.362, 3.669) {};
    \node[disk normal] at (-4.516, 4.233) {};
    \draw[shift={(-1.411, 5.08)}, yscale=1.333, RedOrange, very thick] (0, 0) -- (0, -3.387) -- (0, -3.387);
    \draw[shift={(-0.564, 5.08)}, yscale=1.333, RedOrange, very thick] (0, 0) -- (0, -3.387);
    \draw[very thick] (0, 4.233) -- (-0.564, 3.669) -- (0, 3.104);
    \draw[very thick] (-1.411, 4.233) -- (-1.976, 4.233) -- (-2.258, 4.233);
    \node[disk normal] at (-1.411, 4.233) {};
    \node[disk normal] at (-0.564, 3.669) {};
    \draw[very thick] (-1.411, 4.233) -- (-0.564, 3.669);
    \draw[very thick] (-1.411, 4.233) -- (-0.847, 4.798) -- (-0.847, 4.798);
    \node[anchor=center, font=\LARGE] at (-2.822, 2.822) {$=$};
    \draw[very thick] (-3.951, 2.54) -- (-4.516, 1.976) -- (-3.951, 1.411);
    \draw[very thick] (-4.516, 1.976) -- (-5.362, 1.411) -- (-4.798, 0.847) -- (-4.798, 0.847);
    \draw[very thick] (-5.362, 1.411) -- (-5.927, 1.411) -- (-6.209, 1.411);
    \node[disk normal] at (-5.362, 1.411) {};
    \node[disk normal] at (-4.516, 1.976) {};
    \draw[very thick] (0, 1.976) -- (-0.564, 1.411) -- (0, 0.847);
    \draw[very thick] (-1.411, 1.976) -- (-1.976, 1.976) -- (-2.258, 1.976);
    \node[disk normal] at (-1.411, 1.976) {};
    \node[disk normal] at (-0.564, 1.411) {};
    \draw[very thick] (-1.411, 1.976) -- (-0.564, 1.411);
    \draw[very thick] (-1.411, 1.976) -- (-0.847, 2.54) -- (-0.847, 2.54);
    \node[anchor=center] at (-3.669, 4.798) {$g$};
    \node[anchor=center] at (-3.669, 3.669) {$h$};
    \node[anchor=center] at (-4.647, 3.132) {$k$};
    \node[anchor=center] at (-4.498, 0.453) {$g,h$};
    \node[anchor=center] at (-5.425, 0.416) {$gh,k$};
    \node[anchor=center] at (-0.59, 0.35) {$h,k$};
    \node[anchor=center] at (-1.503, 0.35) {$g,hk$};
\end{tikzpicture} \ ,
\end{equation}
where periodic boundary conditions on the vertical (red) virtual indices are taken.

We now make our most important assumption: that there exist {\it associator} tensors $R_{g,h,k}$, or just {\it associators} for simplicity, together with their left inverses  $\tilde{R}_{g,h,k}$, such that 
\begin{equation}\label{Rinv} R_{g,h,k}\cdot \tilde{R}_{g,h,k} = \id_{g,hk}\otimes \id_{h,k} \ ,
\end{equation}
and such that the following local equation holds between the fusion tensors and associators:
\begin{equation}\label{defR}
\begin{tikzpicture}[scale=1, baseline=0cm]
    \draw[shift={(7.336, 1.816)}, xscale=-0.107, yscale=0.959, RedOrange, very thick] (0, 0) -- (0, -3.387);
    \draw[shift={(6.471, 1.819)}, xscale=4121.39, yscale=0.96, RedOrange, very thick] (0, 0) -- (0, -3.387);
    \draw[shift={(2.539, 1.258)}, xscale=-33.58, yscale=0.8, RedOrange, very thick] (0, 0) -- (0, -3.387);
    \draw[shift={(3.385, 1.278)}, xscale=-33.58, yscale=0.8, RedOrange, very thick] (0, 0) -- (0, -3.387);
    \draw[very thick] (7.902, 0.282) -- (7.338, -0.282) -- (7.902, -0.847);
    \draw[very thick] (6.491, 0.282) -- (5.927, 0.282) -- (5.644, 0.282);
    \node[disk normal] at (6.491, 0.282) {};
    \node[disk normal] at (7.338, -0.282) {};
    \draw[very thick] (6.491, 0.282) -- (7.338, -0.282);
    \draw[very thick] (6.491, 0.282) -- (7.056, 0.847) -- (7.056, 0.847);
    \draw[very thick] (3.951, 0.847) -- (3.387, 0.282) -- (3.951, -0.282);
    \draw[very thick] (3.387, 0.282) -- (2.54, -0.282) -- (3.104, -0.847) -- (3.104, -0.847);
    \draw[very thick] (2.54, -0.282) -- (1.976, -0.282) -- (1.693, -0.282);
    \node[disk normal] at (2.54, -0.282) {};
    \node[disk normal] at (3.387, 0.282) {};
    \node[anchor=center, font=\LARGE] at (4.906, -0.004) {$=$};
    \filldraw[shift={(6.235, 1.645)}, xscale=1.009, yscale=1.19, fill=RedOrange] (0, 0) rectangle (1.401, -0.36);
    \filldraw[shift={(6.203, -0.819)}, xscale=0.997, yscale=1.26, fill=RedOrange] (0, 0) rectangle (1.401, -0.36);
    \node[anchor=center] at (6.916, 1.399) {$R_{g,h,k}$};
    \node[anchor=center] at (3.354, 1.487) {$g,h$};
    \node[anchor=center] at (2.48, 1.475) {$gh,k$};
    \node[anchor=center] at (6.035, 0.916) {$g,hk$};
    \node[anchor=center] at (7.645, 0.881) {$h,k$};
    \node[anchor=center] at (6.916, -1.076) {$\tilde{R}_{g,h,k}$};
\end{tikzpicture}\ .
\end{equation}
Then Eq.~\eqref{assofmpo} follows from the local relation Eq.~\eqref{defR} together with the invertibility assumption Eq.~\eqref{Rinv}.

We note that since the associators $R$ appear together with their inverses in their defining equation \eqref{defR}, they are determined up to a constant factor:
\begin{equation}\label{freedR}
    R(g,h,k)\to \alpha(g,h,k) R(g,h,k)\ ,\ \tilde{R}(g,h,k)\to \bar{\alpha}(g,h,k) \tilde{R}(g,h,k)  \ ,
\end{equation} 
where we have taken $\alpha$ to be a $U(1)$ phase factor without loss of generality so that $\bar{\alpha}$ is its inverse.

{\bf Pentagon relation.} We now consider the concatenation of three fusion MPO tensors. Starting from the configuration $(((g,h) ,k) ,l)$ there are two different ways of repeatedly using Eq.~\eqref{defR} to arrive at $(g,(h,(k,l)))$. These two different ways then give rise to the following equality:

\begin{equation}
\begin{tikzpicture}[scale=1, baseline=2cm]
    \draw[shift={(5.447, 5.099)}, xscale=-151.47, yscale=1.25, RedOrange, very thick] (0, 0) -- (0, -3.387);
    \draw[shift={(6.275, 5.104)}, xscale=2218.59, yscale=1.265, RedOrange, very thick] (0, 0) -- (0, -3.387);
    \draw[shift={(1.691, 5.065)}, xscale=-207.885, yscale=1.124, RedOrange, very thick] (0, 0) -- (0, -3.387);
    \draw[shift={(1.433, 0.351)}, xscale=2699.73, yscale=0.97, RedOrange, very thick] (0, 0) -- (0, -3.387);
    \draw[shift={(2.859, -1.859)}, xscale=0.997, yscale=0.808, RedOrange, very thick] (0, 0) -- (-0.71, -0.284) -- (-0.722, -1.28);
    \filldraw[shift={(1.023, -2.266)}, yscale=1.311, fill=RedOrange] (0, 0) rectangle (1.401, -0.36);
    \draw[shift={(4.949, 0.349)}, xscale=-5325.44, yscale=1.108, RedOrange, very thick] (0, 0) -- (0, -3.387);
    \draw[shift={(6.378, -2.352)}, xscale=0.997, yscale=0.808, RedOrange, very thick] (0, 0) -- (-0.71, -0.284) -- (-0.722, -1.28);
    \draw[shift={(2.141, -0.631)}, xscale=0.982, yscale=0.472, RedOrange, very thick] (0, 0) -- (0.709, 0.364) -- (0.727, 2.025);
    \draw[shift={(2.843, -0.617)}, xscale=0.996, yscale=0.572, RedOrange, very thick] (0, 0) -- (-0.722, 0.313) -- (-0.719, 1.686);
    \draw[shift={(2.841, -0.611)}, xscale=-0.173, yscale=0.368, RedOrange, very thick] (0, 0) -- (0, -3.387);
    \draw[shift={(-0.676, 5.104)}, xscale=0.653, yscale=1.099, RedOrange, very thick] (0, 0) -- (0, -3.387);
    \draw[shift={(-1.517, 5.106)}, xscale=-7203.84, yscale=1.1, RedOrange, very thick] (0, 0) -- (0, -3.387);
    \draw[shift={(-5.096, 4.448)}, xscale=-33.58, yscale=0.842, RedOrange, very thick] (0, 0) -- (0, -3.387);
    \draw[shift={(-4.25, 4.468)}, xscale=-6.715, yscale=0.839, RedOrange, very thick] (0, 0) -- (0, -3.387);
    \draw[very thick] (-0.084, 3.533) -- (-0.648, 2.968) -- (-0.084, 2.404);
    \node[disk normal] at (-1.495, 3.533) {};
    \node[disk normal] at (-0.648, 2.968) {};
    \draw[very thick] (-1.495, 3.533) -- (-0.648, 2.968);
    \draw[very thick] (-1.495, 3.533) -- (-0.931, 4.097) -- (-0.931, 4.097);
    \draw[very thick] (-3.684, 4.036) -- (-4.248, 3.472) -- (-3.684, 2.908);
    \draw[very thick] (-4.248, 3.472) -- (-5.095, 2.908) -- (-4.531, 2.343) -- (-4.531, 2.343);
    \draw[shift={(-5.908, 2.057)}, xscale=0.573, yscale=1.849, very thick] (0, 0) -- (-0.564, 0) -- (-0.847, 0);
    \node[disk normal] at (-5.095, 2.908) {};
    \node[disk normal] at (-4.248, 3.472) {};
    \node[anchor=center, font=\LARGE] at (-3.13, 3.226) {$=$};
    \filldraw[shift={(-1.752, 4.896)}, xscale=1.009, yscale=1.19, fill=RedOrange] (0, 0) rectangle (1.401, -0.36);
    \filldraw[shift={(-1.81, 1.998)}, yscale=1.271, fill=RedOrange] (0, 0) rectangle (1.401, -0.36);
    \node[anchor=center] at (-1.07, 4.649) {$R_{g,h,k}$};
    \node[anchor=center, font=\small] at (-4.281, 4.677) {$g,h$};
    \node[anchor=center, font=\small] at (-5.155, 4.665) {$gh,k$};
    \node[anchor=center] at (-1.096, 1.737) {$\tilde{R}_{g,h,k}$};
    \draw[shift={(-5.933, 4.471)}, xscale=-33.58, yscale=0.839, RedOrange, very thick] (0, 0) -- (0, -3.387);
    \node[disk normal] at (-5.942, 2.061) {};
    \draw[very thick] (-5.08, 2.877) -- (-5.911, 2.063) -- (-5.402, 1.554);
    \node[anchor=center, font=\small] at (-6.112, 4.693) {$ghk,l$};
    \draw[shift={(-2.341, 2.693)}, xscale=0.512, yscale=1.607, very thick] (0, 0) -- (-0.564, 0) -- (-0.847, 0);
    \draw[shift={(-2.365, 5.13)}, xscale=113.973, yscale=1.104, RedOrange, very thick] (0, 0) -- (0, -3.387);
    \node[disk normal] at (-2.375, 2.696) {};
    \draw[very thick] (-1.513, 3.512) -- (-2.344, 2.698) -- (-1.835, 2.189);
    \draw[shift={(3.38, 5.076)}, xscale=0.1, yscale=1.148, RedOrange, very thick] (0, 0) -- (0, -3.387);
    \draw[shift={(2.512, 5.063)}, xscale=5058.92, yscale=1.144, RedOrange, very thick] (0, 0) -- (0, -3.387);
    \node[disk normal] at (2.514, 2.819) {};
    \node[disk normal] at (3.372, 3.117) {};
    \node[anchor=center, font=\LARGE] at (0.507, 3.13) {$=$};
    \filldraw[shift={(2.328, 4.905)}, xscale=1.009, yscale=1.19, fill=RedOrange] (0, 0) rectangle (1.401, -0.36);
    \filldraw[shift={(1.366, 2.378)}, yscale=1.27, fill=RedOrange] (0, 0) rectangle (1.401, -0.36);
    \node[anchor=center] at (3.009, 4.658) {$R_{g,h,k}$};
    \node[anchor=center] at (2.08, 2.118) {$\tilde{R}_{g,hk,l}$};
    \draw[shift={(1.733, 3.354)}, xscale=0.649, yscale=-1.027, very thick] (0, 0) -- (-0.564, 0) -- (-0.847, 0);
    \node[disk normal] at (1.7, 3.357) {};
    \filldraw[shift={(1.457, 4.347)}, xscale=1.009, yscale=1.19, fill=RedOrange] (0, 0) rectangle (1.401, -0.36);
    \node[anchor=center] at (2.139, 4.1) {$R_{g,hk,l}$};
    \filldraw[shift={(2.271, 1.823)}, yscale=1.279, fill=RedOrange] (0, 0) rectangle (1.401, -0.36);
    \node[anchor=center] at (2.984, 1.608) {$\tilde{R}_{g,h,k}$};
    \draw[shift={(7.136, 5.101)}, xscale=0.17, yscale=1.264, RedOrange, very thick] (0, 0) -- (0, -3.387);
    \node[disk normal] at (6.296, 3.043) {};
    \node[disk normal] at (7.172, 2.786) {};
    \filldraw[shift={(6.122, 4.994)}, xscale=1.009, yscale=1.19, fill=RedOrange] (0, 0) rectangle (1.401, -0.36);
    \filldraw[shift={(5.14, 1.958)}, xscale=1.004, yscale=1.264, fill=RedOrange] (0, 0) rectangle (1.401, -0.36);
    \node[anchor=center] at (6.804, 4.748) {$R_{g,h,k}$};
    \node[anchor=center] at (5.86, 1.7) {$\tilde{R}_{g,hk,l}$};
    \draw[shift={(5.52, 3.298)}, xscale=0.649, yscale=-1.027, very thick] (0, 0) -- (-0.564, 0) -- (-0.847, 0);
    \node[disk normal] at (5.481, 3.308) {};
    \filldraw[shift={(5.238, 4.477)}, xscale=1.009, yscale=1.19, fill=RedOrange] (0, 0) rectangle (1.401, -0.36);
    \node[anchor=center] at (5.92, 4.23) {$R_{g,hk,l}$};
    \filldraw[shift={(6.092, 1.423)}, yscale=1.246, fill=RedOrange] (0, 0) rectangle (1.401, -0.36);
    \node[anchor=center] at (6.806, 1.171) {$\tilde{R}_{g,h,k}$};
    \draw[shift={(5.492, 3.325)}, xscale=1.024, yscale=0.642, very thick] (0, 0) -- (1.646, -0.85);
    \draw[shift={(6.272, 3.03)}, xscale=0.71, yscale=0.578, very thick] (0, 0) -- (0.564, 0.564) -- (0.564, 0.564);
    \draw[shift={(5.469, 3.31)}, xscale=0.668, yscale=0.553, very thick] (0, 0) -- (0.564, 0.564) -- (0.564, 0.564);
    \filldraw[shift={(6.107, 3.971)}, xscale=1.009, yscale=1.19, fill=RedOrange] (0, 0) rectangle (1.401, -0.36);
    \node[anchor=center] at (6.789, 3.725) {$R_{h,k,l}$};
    \filldraw[shift={(6.022, 2.511)}, xscale=1.002, yscale=1.259, fill=RedOrange] (0, 0) rectangle (1.401, -0.36);
    \node[anchor=center] at (6.739, 2.255) {$\tilde{R}_{h,k,l}$};
    \node[anchor=center, font=\LARGE] at (4.496, 3.039) {$=$};
    \draw[shift={(-1.603, 0.213)}, xscale=-0.107, yscale=0.959, RedOrange, very thick] (0, 0) -- (0, -3.387);
    \draw[shift={(-2.468, 0.215)}, xscale=4121.39, yscale=0.96, RedOrange, very thick] (0, 0) -- (0, -3.387);
    \node[disk normal] at (-2.443, -1.301) {};
    \node[disk normal] at (-1.598, -1.846) {};
    \draw[very thick] (-2.465, -1.303) -- (-1.618, -1.868);
    \node[anchor=center, font=\LARGE] at (-3.214, -1.413) {$=$};
    \filldraw[shift={(-2.685, 0.056)}, xscale=1.009, yscale=1.19, fill=RedOrange] (0, 0) rectangle (1.401, -0.36);
    \filldraw[shift={(-2.744, -2.189)}, yscale=1.311, fill=RedOrange] (0, 0) rectangle (1.401, -0.36);
    \node[anchor=center] at (-2.003, -0.191) {$R_{gh,k,l}$};
    \node[anchor=center] at (-2.031, -2.464) {$\tilde{R}_{gh,k,l}$};
    \draw[shift={(-2.456, -1.311)}, xscale=0.512, yscale=1.607, very thick] (0, 0) -- (-0.564, 0) -- (-0.847, 0);
    \draw[shift={(-0.722, 0.25)}, xscale=-10.507, yscale=0.965, RedOrange, very thick] (0, 0) -- (0, -3.387);
    \node[disk normal] at (-0.713, -0.639) {};
    \node[anchor=center, font=\LARGE] at (0.507, -1.398) {$=$};
    \draw[very thick] (-1.099, -1.563) -- (-1.584, -1.853) -- (-1.111, -2.199);
    \draw[very thick] (-0.231, -0.35) -- (-0.716, -0.64) -- (-0.243, -0.986);
    \draw[shift={(2.14, -0.609)}, xscale=-0.173, yscale=0.368, RedOrange, very thick] (0, 0) -- (0, -3.387);
    \node[disk normal] at (1.448, -1.37) {};
    \draw[shift={(1.43, -1.372)}, xscale=0.817, yscale=-0.823, very thick] (0, 0) -- (0.847, -0.564);
    \filldraw[shift={(1.037, 0.241)}, xscale=1.009, yscale=1.19, fill=RedOrange] (0, 0) rectangle (1.401, -0.36);
    \node[anchor=center] at (1.74, 0.035) {$R_{gh,k,l}$};
    \node[anchor=center] at (1.746, -2.513) {$\tilde{R}_{gh,k,l}$};
    \draw[shift={(1.435, -1.379)}, xscale=0.512, yscale=1.607, very thick] (0, 0) -- (-0.564, 0) -- (-0.847, 0);
    \node[disk normal] at (2.151, -0.929) {};
    \draw[very thick] (3.328, -1.297) -- (2.843, -1.586) -- (3.317, -1.933);
    \draw[very thick] (2.614, -0.629) -- (2.129, -0.918) -- (2.603, -1.265);
    \node[disk normal] at (2.836, -1.608) {};
    \draw[shift={(2.126, -1.848)}, xscale=0.964, yscale=0.805, RedOrange, very thick] (0, 0) -- (0.689, -0.293) -- (0.681, -1.278) -- (0.681, -1.278);
    \draw[shift={(5.656, -0.633)}, xscale=0.982, yscale=0.472, RedOrange, very thick] (0, 0) -- (0.709, 0.364) -- (0.727, 2.025);
    \draw[shift={(6.358, -0.619)}, xscale=0.996, yscale=0.572, RedOrange, very thick] (0, 0) -- (-0.722, 0.313) -- (-0.719, 1.686);
    \draw[shift={(6.356, -0.614)}, xscale=-1.964, yscale=0.515, RedOrange, very thick] (0, 0) -- (0, -3.387);
    \node[anchor=center, font=\LARGE] at (4.022, -1.4) {$=$};
    \draw[shift={(5.655, -0.611)}, xscale=-2.757, yscale=0.511, RedOrange, very thick] (0, 0) -- (0, -3.387);
    \node[disk normal] at (4.954, -1.4) {};
    \filldraw[shift={(4.655, 0.241)}, xscale=0.856, yscale=1.195, fill=RedOrange] (0, 0) rectangle (1.401, -0.36);
    \filldraw[shift={(4.649, -2.762)}, xscale=0.86, yscale=1.299, fill=RedOrange] (0, 0) rectangle (1.401, -0.36);
    \node[anchor=center] at (5.255, 0.033) {$R_{gh,k,l}$};
    \node[anchor=center] at (5.248, -3.004) {$\tilde{R}_{gh,k,l}$};
    \draw[shift={(4.965, -1.421)}, xscale=0.512, yscale=1.607, very thick] (0, 0) -- (-0.564, 0) -- (-0.847, 0);
    \node[disk normal] at (5.681, -1.606) {};
    \draw[very thick] (6.861, -1.504) -- (6.376, -1.793) -- (6.85, -2.14);
    \node[disk normal] at (6.365, -1.798) {};
    \draw[shift={(5.644, -2.34)}, xscale=0.964, yscale=0.805, RedOrange, very thick] (0, 0) -- (0.689, -0.293) -- (0.681, -1.278) -- (0.681, -1.278);
    \draw[very thick] (4.971, -1.403) -- (5.326, -1.173);
    \draw[very thick] (5.697, -1.592) -- (6.025, -1.376);
    \filldraw[shift={(4.666, -0.704)}, xscale=0.856, yscale=1.195, fill=RedOrange] (0, 0) rectangle (1.401, -0.36);
    \node[anchor=center] at (5.343, -0.937) {$R_{g,h,kl}$};
    \filldraw[shift={(4.646, -1.763)}, xscale=0.86, yscale=1.299, fill=RedOrange] (0, 0) rectangle (1.401, -0.36);
    \node[anchor=center] at (5.245, -2.005) {$\tilde{R}_{g,h,kl}$};
    \node[anchor=center] at (-3.531, 4.038) {$g$};
    \node[anchor=center] at (-3.58, 2.918) {$h$};
    \node[anchor=center] at (-4.419, 2.329) {$k$};
    \node[anchor=center] at (-5.278, 1.528) {$l$};
    \draw[very thick] (7.681, 3.073) -- (7.196, 2.783) -- (7.67, 2.436);
    \draw[very thick] (1.705, 3.357) -- (2.502, 2.812) -- (3.361, 3.126) -- (3.781, 3.42);
    \draw[very thick] (3.39, 3.12) -- (3.776, 2.873);
    \draw[shift={(2.535, 2.798)}, xscale=1.443, yscale=1.325, very thick] (0, 0) -- (0.259, -0.199);
    \draw[very thick] (1.726, 3.374) -- (2.14, 3.689);
    \draw[very thick] (-2.432, -1.304) -- (-0.731, -0.645);
    \draw[very thick] (1.453, -1.387) -- (2.827, -1.588);
    \draw[very thick] (4.949, -1.405) -- (6.376, -1.807);
\end{tikzpicture}
\end{equation}

From this equation we now apply the orthogonality relations Eq.~\eqref{orthoF} and Eq.~\eqref{Rinv} to invert the $F$ and the $R$ tensors to conclude that the following configuration of associators is proportional to the identity, where we dub such constant $\omega:$

\begin{equation}\label{defomega}
\begin{tikzpicture}[scale=1]
    \draw[RedOrange, very thick] (-5.568, 3.877) -- (2.068, 3.901);
    \draw[RedOrange, very thick] (-5.567, 3.037) -- (0.043, 3.053) -- (0.898, 2.215) -- (2.033, 2.21);
    \draw[RedOrange, very thick] (0.509, 2.662) -- (0.862, 3.024) -- (2.04, 3.019);
    \draw[RedOrange, very thick] (-5.567, 2.184) -- (0.086, 2.207) -- (0.443, 2.588);
    \node[anchor=center] at (-1.168, 2.408) {$g,h$};
    \node[anchor=center] at (-1.361, 3.235) {$gh,k$};
    \node[anchor=center] at (-2.913, 3.241) {$g,hk$};
    \node[anchor=center] at (-3.437, 2.409) {$h,k$};
    \filldraw[shift={(-0.827, 2.803)}, rotate=89.772, fill=RedOrange] (0, 0) rectangle (1.401, -0.36);
    \node[anchor=center] at (-5.872, 2.211) {$k,l$};
    \node[anchor=center] at (-5.963, 3.024) {$h,kl$};
    \node[anchor=center] at (-6.049, 3.895) {$g,hkl$};
    \node[anchor=center] at (-4.332, 3.2) {$hk,l$};
    \node[anchor=center] at (-2.212, 4.148) {$ghk,l$};
    \node[anchor=center] at (2.281, 2.207) {$k,l$};
    \node[anchor=center] at (0.487, 4.105) {$gh,kl$};
    \filldraw[shift={(1.354, 2.815)}, rotate=89.772, fill=RedOrange] (0, 0) rectangle (1.401, -0.36);
    \node[anchor=center] at (2.399, 3.025) {$h,kl$};
    \node[anchor=center] at (2.501, 3.944) {$g,hkl$};
    \node[anchor=center] at (5.7, 3.164) {$= \omega(g,h,k,l) \cdot \id_{k,l} \otimes \id_{h,kl}\otimes \id_{g,hkl}$};
    \filldraw[shift={(-2.397, 1.896)}, rotate=89.772, fill=RedOrange] (0, 0) rectangle (1.401, -0.36);
    \filldraw[shift={(-5.152, 1.908)}, rotate=89.772, fill=RedOrange] (0, 0) rectangle (1.401, -0.36);
    \filldraw[shift={(-3.829, 2.796)}, rotate=89.772, fill=RedOrange] (0, 0) rectangle (1.401, -0.36);
\end{tikzpicture}
\end{equation}

This structure has a very natural geometric interpretation: it is a 5-cell (4-simplex) whose projection to 3D onto a triangular bipyramid, with an extra edge joining the top and bottom vertex, allow us to compute $\omega$ in a closed form: 
$$
\begin{tikzpicture}[scale=1]
    \draw[RedOrange, thick] (-3.669, 2.258) -- (0.847, 2.258) -- (-2.54, 1.129) -- (-3.669, 2.258);
    \draw[RedOrange, thick] (-3.669, 2.258) -- (-1.411, 4.516) -- (-2.54, 1.129) -- (-1.411, -1.129) -- (-1.411, 4.516) -- (0.847, 2.258) -- (-1.411, -1.129) -- (-3.669, 2.258);
    \node[fsquare large, fill=BrickRed!90] at (-1.411, 4.516) {};
    \node[fsquare large, fill=BrickRed!90] at (0.847, 2.258) {};
    \node[fsquare large, fill=BrickRed!90] at (-2.54, 1.129) {};
    \node[fsquare large, fill=BrickRed!90] at (-3.669, 2.258) {};
    \node[fsquare large, fill=BrickRed!90] at (-1.411, -1.129) {};
    \draw[shift={(-1.478, 1.522)}, xscale=0.956, yscale=1.387, white, very thick] (0, 0) -- (0.155, 0.036) -- (0.155, 0.036);
    \draw[shift={(-1.462, 1.451)}, xscale=0.945, yscale=1.326, white, very thick] (0, 0) -- (0.155, 0.036) -- (0.155, 0.036);
    \draw[white, very thick] (-1.447, 2.4) -- (-1.447, 2.118);
    \draw[white, very thick] (-1.375, 2.394) -- (-1.375, 2.111);
    \draw[white, very thick] (-2.08, 2.395) -- (-2.182, 2.092);
    \draw[white, very thick] (-2.147, 2.416) -- (-2.249, 2.112);
    \node[anchor=center] at (-2.756, 3.813) {$g,hkl$};
    \node[anchor=center] at (-2.162, 2.92) {$hk,l$};
    \node[anchor=center] at (-0.993, 2.878) {$ghk,l$};
    \node[anchor=center] at (0.235, 3.448) {$g,hk$};
    \node[anchor=center] at (-0.722, 1.412) {$h,k$};
    \node[anchor=center] at (0.261, 0.572) {$gh,k$};
    \node[anchor=center] at (-2.744, 2.342) {$g,h$};
    \node[anchor=center] at (-1.96, 0.531) {$k,l$};
    \node[anchor=center] at (-2.808, 1.73) {$h,kl$};
    \node[anchor=center] at (-2.861, 0.266) {$gh,kl$};
    \node[anchor=center] at (-1.405, 4.843) {$R(g,hk,l)$};
    \node[anchor=center] at (-0.364, -1.088) {$\tilde{R}(gh,k,l)$};
    \node[anchor=center] at (1.482, 2.585) {$R(g,h,k)$};
    \node[anchor=center] at (-4.405, 2.596) {$\tilde{R}(g,h,kl)$};
    \node[anchor=center] at (-3.591, 0.954) {$R(h,k,l)$};
    \node[anchor=center, font=\large] at (4.128, 2.148) {$= \omega(g,h,k,l) \cdot D$};
\end{tikzpicture}
\ ,
$$
where $D$ is the product of the dimensions of the virtual spaces ${(k,l)}$, ${(h,kl)}$ and  ${(g,hkl)}$.

The freedom in defining $R$ up to a constant, Eq.~\eqref{freedR}, translates into the following freedom in the  definition of $\omega$:
\begin{equation}\label{freedomega}
\omega(g, h, k, l)\to \frac{\alpha(g,h,k)\alpha(g,hk,l)\alpha(h,k,l)}{\alpha(gh,k,l)\alpha(g,h,kl)}\omega(g, h, k, l )
\end{equation}

{\bf Associahedron relation.} The different compatible  ways to modify the concatenation of four fusion MPO tensors enforces the $\omega$ factors to satisfy the $4$-cocycle equation:
\begin{equation}\label{4cocycleq}
\omega(g,h, k, l m)\omega(g, h k, l, m) \omega(h,k, l, m)=\omega(g, h, k l, m) \omega(g h,k, l, m)\omega(g, h, k, l) \ , 
\end{equation}
whose solutions up to the equivalence relation
Eq.~\eqref{freedomega} are classified by the cohomology group $H^4(G,U(1))$. The full associahedron relation can be checked in Fig.18 of Ref.\cite{Kitaev06} in terms of the fusion spaces.

\subsection{Triple line realization}

In this section we construct a particular realization of the tensors from the previous section. This realization is the simplest example for the above relations, however, it fails to be unitary: we will see that $O_1 \neq \id $, instead, it is a  projector; similarly, $O_g O_g^\dagger = O_g O_{g^{-1}} = O_1$, so $O_g$ is not unitary on the full space but only on the subspace projecte by $O_1$.   

The construction closely follows ref.~\cite{Bultinck17A} that provides examples for the one-dimensional MPO case based a $3$-cocycle of a finite group. As our examples are in 2D, we use a $4$-cocycle instead of a $3$-cocycle. As each index of the tensor network is broken into a tensor product of three smaller indices, the construction is called triple-line representation.

We first notice that the Eqs.~\eqref{prodTgh}, \eqref{defR} and \eqref{defomega} always involves six tensors among $T_g$, $F_{g,h}$ and $ R_{g,h,k}$. Crucially,  the $4$-cocycle equation, \eqref{4cocycleq}, is a relation between six $\omega's$. As such, we will use it to assign appropriate $\omega$ values to each tensors $T_g$, $F_{g,h}$ and $ R_{g,h,k}$ in such a way that those equations will reproduce the 4-cocycle equation \eqref{4cocycleq} and thus they will automatically be  satisfied.

The tensors of the TN operator representing the symmetry is given by the following non-zero contributions:
$$
\begin{tikzpicture}[scale=1]
    \draw[very thick] (-4.117, 5.435) .. controls (-4.085, 4.433) and (-3.865, 4.373) .. (-2.834, 4.403);
    \draw[very thick] (-4.385, 5.378) .. controls (-4.362, 4.656) and (-4.549, 4.547) .. (-5.363, 5.251);
    \draw[very thick, thick bevel] (-4.268, 5.458) .. controls (-4.259, 4.737) and (-4.258, 4.571) .. (-4.519, 4.236) .. controls (-4.78, 3.902) and (-5.302, 3.399) .. (-5.302, 3.399);
    \draw[very thick] (-5.423, 4.998) .. controls (-4.664, 4.357) and (-4.486, 4.259) .. (-4.422, 3.903) .. controls (-4.359, 3.546) and (-4.41, 2.929) .. (-4.41, 2.929);
    \draw[very thick] (-2.8, 4.173) .. controls (-4.037, 4.185) and (-4.162, 4.081) .. (-4.031, 2.833);
    \node[anchor=center, font=\small] at (-5.164, 4.449) {$k_1$};
    \node[anchor=center, font=\small] at (-4.839, 3.125) {$k_2$};
    \node[anchor=center, font=\small] at (-3.586, 3.83) {$k_3$};
    \node[anchor=center, font=\small] at (-3.466, 4.663) {$gk_3$};
    \node[anchor=center, font=\small] at (-4.739, 5.189) {$gk_1$};
    \node[anchor=center, font=\small] at (-5.354, 3.765) {$gk_2$};
    \node[anchor=center, font=\small] at (-1.007, 4.315) {$= \omega(g,k_1,k_1^{-1}k_2,k_2^{-1}k_3)$};
    \node[anchor=center] at (-6.237, 4.335) {$T_g = $};
    \draw[very thick, thick bevel] (-5.144, 3.318) .. controls (-4.196, 4.536) and (-4.077, 4.534) .. (-4.247, 2.856);
    \draw[black!50, densely dashed] (-5.203, 3.394) -- (-4.243, 4.442);
    \draw[black!50, densely dashed] (-5.386, 5.104) -- (-4.253, 4.413) -- (-2.846, 4.277);
    \node[anchor=center, font=\small, text=black!50] at (-4.106, 4.281) {$g$};
\end{tikzpicture} 
$$
$$
\begin{tikzpicture}[scale=1]
    \draw[black!50, densely dashed] (-1.871, 1.22) -- (-0.307, 1.21);
    \draw[black!50, densely dashed] (0.47, 0.027) -- (-0.337, 1.244);
    \draw[black!50, densely dashed] (1.077, 2.372) -- (-0.346, 1.241);
    \draw[very thick] (-0.068, 2.601) .. controls (-0.083, 1.955) and (-0.067, 1.777) .. (-0.028, 1.703) .. controls (0.012, 1.63) and (0.076, 1.661) .. (1.026, 2.425);
    \draw[very thick] (1.069, 2.291) .. controls (0.024, 1.485) and (-0.02, 1.418) .. (-0.034, 1.337) .. controls (-0.049, 1.257) and (-0.033, 1.163) .. (-0.063, -0.25);
    \draw[very thick] (-0.351, 2.474) .. controls (-0.33, 1.63) and (-0.353, 1.5) .. (-0.366, 1.399) .. controls (-0.38, 1.298) and (-0.385, 1.227) .. (-1.848, 1.272);
    \draw[very thick] (-1.836, 1.156) .. controls (-0.754, 1.139) and (-0.609, 1.149) .. (-0.513, 1.12) .. controls (-0.416, 1.092) and (-0.368, 1.024) .. (-0.43, -0.327);
    \draw[very thick, thick bevel] (-0.259, -0.346) .. controls (-0.268, 1.005) and (-0.233, 0.989) .. (-0.204, 0.957) .. controls (-0.174, 0.925) and (-0.15, 0.878) .. (0.385, 0.004);
    \draw[very thick, thick bevel] (-0.198, 2.552) .. controls (-0.211, 1.386) and (-0.211, 1.304) .. (-0.173, 1.185) .. controls (-0.135, 1.065) and (-0.059, 0.908) .. (0.502, 0.108);
    \node[anchor=center, font=\small] at (0.3, 1.378) {$k_2$};
    \node[anchor=center, font=\small] at (0.28, 2.251) {$gk_2$};
    \node[anchor=center, font=\small] at (-0.696, 0.865) {$k_1$};
    \node[anchor=center, font=\small] at (-0.681, 1.627) {$gk_1$};
    \node[anchor=center, font=\small] at (0.241, -0.048) {$k_3$};
    \node[anchor=center, font=\small] at (0.529, 0.652) {$gk_3$};
    \node[anchor=center, font=\footnotesize, text=black!50] at (-0.334, 1.102) {$g$};
    \node[anchor=center] at (-2.514, 1.199) {$T'_g=$};
    \node[anchor=center] at (3.111, 1.233) {$= \bar{\omega}(g,k_1,k_1^{-1}k_2,k_2^{-1}k_3)$};
\end{tikzpicture}
$$

As mentioned above, these TN operators are not unitaries, since $O_1$ is a non-trivial projector. Assuming that $\omega$ is a normalized $4$-cocycle, i.e.,  $\omega(1,\cdot,\cdot,\cdot)=1$, the tensor $T_1$ is simply a tensor product of maximally entangled states, and thus $O_1$ projects onto the state formed by taking the tensor product of  maximally entangled states $\ket{\omega} = \sum_g \ket{gg}$ on every edge: $O_1 = \ket{\Psi}\bra{\Psi}$, with $\ket{\Psi} = \bigotimes_{e: \text{ edge}} \ket{\omega_1}$,

\begin{equation}\label{3Lview}
\ket{\Psi} = 
\begin{tikzpicture}[scale=1,baseline=-1.3cm]
    \draw[shift={(-1.134, 0.881)}, scale=0.687, very thick] (0, 0) -- (0.002, -0.718);
    \draw[shift={(-0.98, 0.13)}, scale=0.925, very thick] (0, 0) -- (0.87, -0.87);
    \draw[very thick] (-1.274, 0.16) -- (-2.129, -0.695);
    \draw[very thick] (-2.271, -0.974) -- (-2.271, -1.809);
    \draw[very thick] (-2.121, -2.116) -- (-1.312, -2.925);
    \draw[very thick] (-0.923, -2.876) -- (-0.129, -2.081);
    \draw[very thick] (-0.013, -1.794) -- (-0.013, -1.015);
    \draw[very thick] (0.127, -0.697) -- (0.639, -0.184);
    \draw[shift={(0.113, -2.092)}, scale=0.801, very thick] (0, 0) -- (0.565, -0.565);
    \draw[very thick] (-2.406, -0.702) -- (-2.837, -0.271);
    \draw[shift={(-2.418, -2.112)}, scale=0.916, very thick] (0, 0) -- (-0.522, -0.522);
    \draw[very thick] (-1.129, -3.104) -- (-1.129, -3.669);
    \node[disk normal] at (-2.425, -0.683) {};
    \node[disk normal] at (-2.271, -0.974) {};
    \node[disk normal] at (-2.125, -0.69) {};
    \node[disk normal] at (-2.42, -2.115) {};
    \node[disk normal] at (-2.121, -2.116) {};
    \node[disk normal] at (-2.271, -1.809) {};
    \node[disk normal] at (-0.138, -2.091) {};
    \node[disk normal] at (-0.013, -1.778) {};
    \node[disk normal] at (0.125, -2.104) {};
    \node[disk normal] at (-0.917, -2.87) {};
    \node[disk normal] at (-1.312, -2.925) {};
    \node[disk normal] at (-1.129, -3.111) {};
    \node[disk normal] at (-0.175, -0.675) {};
    \node[disk normal] at (-0.013, -1.015) {};
    \node[disk normal] at (0.127, -0.697) {};
    \node[disk normal] at (-0.998, 0.148) {};
    \node[disk normal] at (-1.287, 0.147) {};
    \node[disk normal] at (-1.132, 0.405) {};
\end{tikzpicture}
\end{equation}

The values of the fusion MPO tensors in Eq.~\eqref{prodTgh} are assigned as follows:
$$
\begin{tikzpicture}[scale=1]
    \draw[very thick] (3.513, 1.069) .. controls (3.544, 0.067) and (3.764, 0.007) .. (4.795, 0.037);
    \draw[very thick] (3.244, 1.012) .. controls (3.268, 0.29) and (3.081, 0.181) .. (2.267, 0.884);
    \draw[very thick, thick bevel] (3.362, 1.092) .. controls (3.371, 0.371) and (3.371, 0.204) .. (3.111, -0.13) .. controls (2.85, -0.464) and (2.328, -0.967) .. (2.328, -0.967);
    \draw[very thick] (2.207, 0.631) .. controls (2.966, -0.01) and (3.144, -0.107) .. (3.207, -0.464) .. controls (3.271, -0.821) and (3.22, -1.437) .. (3.22, -1.437);
    \draw[very thick] (4.83, -0.193) .. controls (3.593, -0.181) and (3.468, -0.286) .. (3.599, -1.534);
    \node[anchor=center, font=\small] at (2.465, 0.083) {$k_1$};
    \node[anchor=center, font=\small] at (2.791, -1.241) {$k_2$};
    \node[anchor=center, font=\small] at (4.044, -0.537) {$k_3$};
    \node[anchor=center, font=\small] at (4.164, 0.297) {$ghk_3$};
    \node[anchor=center, font=\small] at (2.891, 0.823) {$ghk_1$};
    \node[anchor=center, font=\small] at (2.276, -0.602) {$ghk_2$};
    \draw[very thick, thick bevel] (2.485, -1.048) .. controls (3.434, 0.169) and (3.553, 0.167) .. (3.383, -1.51);
    \draw[very thick] (4.792, 0.027) .. controls (5.07, 0.152) and (5.011, 0.4) .. (5.013, 0.55) .. controls (5.016, 0.7) and (5.08, 0.753) .. (5.537, 0.772);
    \draw[very thick] (4.82, -0.195) .. controls (5.045, -0.364) and (5.017, -0.52) .. (5.071, -0.609) .. controls (5.126, -0.698) and (5.263, -0.721) .. (5.672, -0.686);
    \draw[very thick] (5.638, 0.396) .. controls (5.201, 0.264) and (5.193, 0.114) .. (5.247, 0.003) .. controls (5.301, -0.109) and (5.416, -0.181) .. (5.702, -0.143);
    \draw[very thick] (1.413, -0.735) .. controls (1.783, -0.219) and (1.881, -0.285) .. (1.936, -0.43) .. controls (1.991, -0.575) and (2.004, -0.799) .. (2.034, -0.936) .. controls (2.065, -1.072) and (2.113, -1.119) .. (2.181, -1.087) .. controls (2.249, -1.056) and (2.336, -0.944) .. (2.336, -0.944);
    \draw[very thick] (2.504, -1.025) .. controls (2.202, -1.247) and (2.209, -1.311) .. (2.256, -1.423) .. controls (2.303, -1.534) and (2.39, -1.695) .. (1.942, -1.922);
    \draw[very thick] (1.482, -1.047) .. controls (1.751, -0.913) and (1.87, -1.056) .. (1.923, -1.171) .. controls (1.977, -1.285) and (1.966, -1.371) .. (1.9, -1.434) .. controls (1.834, -1.497) and (1.713, -1.538) .. (1.713, -1.538);
    \draw[very thick] (2.22, 0.614) .. controls (1.984, 0.877) and (1.93, 0.747) .. (1.912, 0.631) .. controls (1.895, 0.516) and (1.913, 0.416) .. (1.882, 0.367) .. controls (1.852, 0.317) and (1.772, 0.319) .. (1.639, 0.435) .. controls (1.505, 0.551) and (1.316, 0.781) .. (1.352, 0.781);
    \draw[very thick] (2.285, 0.869) .. controls (2.067, 0.969) and (2.047, 1.075) .. (2.079, 1.171) .. controls (2.111, 1.266) and (2.194, 1.352) .. (2.1, 1.484) .. controls (2.006, 1.616) and (1.735, 1.795) .. (1.735, 1.795);
    \draw[very thick] (1.644, 1.446) .. controls (1.828, 1.294) and (1.828, 1.15) .. (1.79, 1.049) .. controls (1.751, 0.948) and (1.675, 0.89) .. (1.432, 1.112);
    \draw[very thick] (-1.854, 1.687) .. controls (-1.877, 1.201) and (-1.948, 1.225) .. (-2.162, 1.321) .. controls (-2.377, 1.417) and (-2.735, 1.586) .. (-2.735, 1.586);
    \draw[very thick] (-1.652, 1.664) .. controls (-1.626, 1.129) and (-1.402, 1.093) .. (-0.493, 1.136);
    \draw[very thick] (-2.944, 1.273) .. controls (-1.973, 0.747) and (-1.967, 0.358) .. (-1.966, 0.03) .. controls (-1.965, -0.298) and (-1.97, -0.564) .. (-1.994, -0.698) .. controls (-2.019, -0.832) and (-2.063, -0.833) .. (-2.842, -0.403);
    \draw[very thick] (-0.403, 0.853) .. controls (-1.304, 0.855) and (-1.414, 0.766) .. (-1.466, 0.612) .. controls (-1.519, 0.459) and (-1.515, 0.24) .. (-1.511, 0.004) .. controls (-1.508, -0.232) and (-1.505, -0.485) .. (-1.452, -0.65) .. controls (-1.4, -0.816) and (-1.297, -0.894) .. (-0.368, -0.859);
    \draw[very thick, thick bevel] (-1.783, 1.546) .. controls (-1.783, 0.994) and (-1.855, 0.724) .. (-2.471, 0.069);
    \draw[very thick] (-1.627, -2.117) .. controls (-1.666, -1.294) and (-1.598, -1.232) .. (-1.48, -1.199) .. controls (-1.363, -1.165) and (-1.197, -1.16) .. (-0.515, -1.181);
    \draw[very thick] (-2.863, -0.698) .. controls (-1.935, -1.159) and (-1.927, -1.504) .. (-1.948, -2.521);
    \draw[very thick, thick bevel] (-2.345, -0.187) .. controls (-1.735, 0.75) and (-1.721, 0.666) .. (-1.719, 0.393) .. controls (-1.718, 0.119) and (-1.729, -0.345) .. (-1.728, -0.573) .. controls (-1.728, -0.801) and (-1.715, -0.792) .. (-2.428, -1.872);
    \draw[very thick, thick bevel] (-2.39, -2.186) .. controls (-1.823, -1.27) and (-1.787, -1.25) .. (-1.767, -1.266) .. controls (-1.748, -1.282) and (-1.746, -1.336) .. (-1.745, -2.407);
    \draw[RedOrange, thick, dotted] (3.381, -0.149) circle[radius=1.702];
    \node[anchor=center, font=\small] at (3.91, -2.032) {$k_3$};
    \node[anchor=center, font=\small] at (4.141, 1.665) {$k_1$};
    \node[anchor=center, font=\small] at (1.187, 0.066) {$k_2$};
    \node[anchor=center, font=\small] at (5.692, 0.095) {$hk_3$};
    \node[anchor=center, font=\small] at (1.304, 1.239) {$hk_1$};
    \node[anchor=center, font=\small] at (1.453, -1.269) {$hk_2$};
    \node[anchor=center, font=\footnotesize] at (6.213, 1.146) {$\bar{\omega}(g,h,k_1,k_1^{-1}k_3)$};
    \node[anchor=center, font=\footnotesize] at (1.335, 2.052) {${\omega}(g,h,k_1,k_1^{-1}k_2)$};
    \node[anchor=center, font=\footnotesize] at (1.809, -2.149) {${\omega}(g,h,k_2,k_2^{-1}k_3)$};
    \node[anchor=center, font=\large] at (0.032, 0.018) {$=$};
    \node[anchor=center, font=\small] at (-1.123, -0.027) {$hk_3$};
    \node[anchor=center, font=\small] at (-1.099, -1.477) {$k_3$};
    \node[anchor=center, font=\small] at (-0.934, 1.44) {$ghk_3$};
    \node[anchor=center, font=\small] at (-2.274, 1.728) {$ghk_1$};
    \node[anchor=center, font=\small] at (-2.711, 0.797) {$hk_1$};
    \node[anchor=center, font=\small] at (-2.557, -1.126) {$k_1$};
    \node[anchor=center, font=\small] at (-2.188, 1.071) {$ghk_2$};
    \node[anchor=center, font=\small] at (-2.016, -0.509) {$hk_2$};
    \node[anchor=center, font=\small] at (-2.309, -2.344) {$k_2$};
    \draw[black!50, densely dashed] (-2.405, -0.048) -- (-1.646, 0.93) -- (-0.447, 0.998);
    \draw[black!50, densely dashed] (-1.633, 0.931) -- (-2.839, 1.421);
    \draw[black!50, densely dashed] (-2.467, -2.058) -- (-1.69, -0.987) -- (-0.389, -1.029);
    \draw[black!50, densely dashed] (-1.696, -0.994) -- (-2.83, -0.615);
    \node[anchor=center, font=\small, text=black!50] at (-1.581, 0.774) {$g$};
    \node[anchor=center, font=\small, text=black!50] at (-1.581, -0.824) {$h$};
    \draw[black!50, densely dashed] (1.649, 1.68) -- (1.957, 1.383) -- (1.785, 0.535) -- (1.335, 0.981);
    \draw[black!50, densely dashed] (1.473, -0.848) -- (1.789, -0.628) -- (2.151, -1.486) -- (1.838, -1.766);
    \draw[black!50, densely dashed] (5.636, 0.637) -- (5.171, 0.441) -- (5.119, -0.418) -- (5.785, -0.419);
    \draw[black!50, densely dashed] (3.506, -0.04) -- (5.137, -0.055);
    \draw[black!50, densely dashed] (3.513, -0.04) -- (1.91, 0.937);
    \draw[black!50, densely dashed] (2.114, -1.296) -- (3.388, 0.056);
    \node[anchor=center, font=\small, text=black!50] at (1.557, 1.677) {$g$};
    \node[anchor=center, font=\small, text=black!50] at (1.354, -0.882) {$g$};
    \node[anchor=center, font=\small, text=black!50] at (5.746, 0.615) {$g$};
    \node[anchor=center, font=\small, text=black!50] at (1.28, 0.975) {$h$};
    \node[anchor=center, font=\small, text=black!50] at (1.753, -1.763) {$h$};
    \node[anchor=center, font=\small, text=black!50] at (5.918, -0.424) {$h$};
    \node[anchor=center, font=\small, text=black!50] at (3.571, -0.183) {$gh$};
\end{tikzpicture}
\ ,
$$
where the red indices carry a label $g,h$ and they have an internal state, with bond dimension $|G|$, indexed by $k_i$ which is the only one depicted for simplicity. From their values it is clear that they satisfy the orthogonality relations \eqref{orthoF} and thus $O_g O_h= O_{gh}$ is satisfied.

The values of the associators are given in Eq.\eqref{defR} by: 
$$
\begin{tikzpicture}[scale=1]
    \draw[shift={(3.42, -0.972)}, xscale=0.901, yscale=1.432, RedOrange, thick, densely dotted] (0, 0) -- (-0.001, 3.613);
    \filldraw[thick, fill=RedOrange] (3.594, 3.378) rectangle (1.539, 3.885);
    \draw[shift={(1.862, 4.169)}, xscale=1.014, yscale=1.413, RedOrange, thick, densely dotted] (0, 0) -- (0, -3.671);
    \draw[RedOrange, thick, densely dotted] (-3.987, 3.274) -- (-3.98, -0.299);
    \draw[RedOrange, thick, densely dotted] (-2.4, 3.301) -- (-2.38, -0.473);
    \draw[thick] (-1.534, 2.743) .. controls (-2.527, 2.749) and (-2.541, 2.597) .. (-2.548, 2.48) .. controls (-2.554, 2.364) and (-2.554, 2.285) .. (-2.559, 2.217) .. controls (-2.564, 2.148) and (-2.574, 2.092) .. (-2.803, 2.064) .. controls (-3.032, 2.036) and (-3.48, 2.038) .. (-3.748, 2.013) .. controls (-4.016, 1.988) and (-4.105, 1.937) .. (-4.149, 1.846) .. controls (-4.194, 1.755) and (-4.194, 1.624) .. (-4.259, 1.499) .. controls (-4.325, 1.374) and (-4.456, 1.254) .. (-5.418, 1.342);
    \draw[thick] (-1.469, 1.358) .. controls (-2.144, 1.346) and (-2.287, 1.548) .. (-2.366, 1.664) .. controls (-2.444, 1.78) and (-2.458, 1.81) .. (-2.67, 1.794) .. controls (-2.882, 1.779) and (-3.292, 1.717) .. (-3.517, 1.643) .. controls (-3.741, 1.569) and (-3.779, 1.483) .. (-3.793, 1.321) .. controls (-3.807, 1.16) and (-3.796, 0.922) .. (-3.229, 0.903);
    \draw[thick] (-3.24, 0.513) .. controls (-4.119, 0.582) and (-4.133, 0.715) .. (-4.143, 0.866) .. controls (-4.154, 1.016) and (-4.162, 1.184) .. (-5.427, 1.053);
    \draw[thick] (-1.496, 2.426) .. controls (-2.145, 2.315) and (-2.173, 2.16) .. (-2.137, 2.027) .. controls (-2.102, 1.895) and (-2.003, 1.785) .. (-1.477, 1.845);
    \node[anchor=center, font=\Large] at (-0.304, 1.764) {$=$};
    \draw[thick] (0.965, 1.675) .. controls (1.49, 1.669) and (1.591, 1.584) .. (1.638, 1.491) .. controls (1.685, 1.398) and (1.678, 1.296) .. (1.727, 1.213) .. controls (1.775, 1.129) and (1.879, 1.064) .. (2.106, 1.043) .. controls (2.332, 1.021) and (2.683, 1.044) .. (2.902, 1.008) .. controls (3.122, 0.972) and (3.211, 0.877) .. (3.281, 0.738) .. controls (3.351, 0.6) and (3.403, 0.418) .. (4.15, 0.432);
    \draw[thick] (4.127, 0.772) .. controls (3.776, 0.78) and (3.672, 0.877) .. (3.63, 1.001) .. controls (3.587, 1.126) and (3.606, 1.278) .. (4.111, 1.341);
    \draw[thick] (4.063, 1.757) .. controls (3.516, 1.666) and (3.439, 1.533) .. (3.377, 1.452) .. controls (3.316, 1.372) and (3.271, 1.345) .. (3.083, 1.351) .. controls (2.894, 1.357) and (2.562, 1.397) .. (2.366, 1.449) .. controls (2.169, 1.501) and (2.108, 1.566) .. (2.075, 1.654) .. controls (2.042, 1.741) and (2.036, 1.851) .. (2.042, 1.942) .. controls (2.048, 2.033) and (2.066, 2.106) .. (2.134, 2.145) .. controls (2.201, 2.184) and (2.319, 2.188) .. (2.577, 2.198);
    \draw[thick] (2.58, 2.729) .. controls (1.96, 2.685) and (1.879, 2.582) .. (1.83, 2.464) .. controls (1.781, 2.347) and (1.763, 2.214) .. (1.747, 2.121) .. controls (1.732, 2.029) and (1.718, 1.977) .. (0.949, 2.016);
    \draw[black!50, thick, densely dashed] (-5.421, 1.234) -- (-4.016, 1.248) -- (-3.912, 0.835) -- (-3.155, 0.721);
    \draw[black!50, thick, densely dashed] (-4.015, 1.246) -- (-3.993, 1.805) -- (-2.348, 1.97) -- (-2.368, 2.421) -- (-1.495, 2.574);
    \draw[black!50, thick, densely dashed] (-2.349, 1.978) -- (-2.068, 1.651) -- (-1.496, 1.61);
    \draw[black!50, thick, densely dashed] (2.57, 2.491) -- (2.021, 2.4) -- (1.854, 1.77) -- (0.938, 1.86);
    \draw[black!50, thick, densely dashed] (1.854, 1.77) -- (1.94, 1.337) -- (3.408, 1.104) -- (3.567, 1.399) -- (4.056, 1.567);
    \draw[black!50, thick, densely dashed] (3.406, 1.105) -- (3.532, 0.72) -- (4.097, 0.587);
    \node[anchor=center] at (-4.409, 0.913) {$t$};
    \node[anchor=center] at (2.602, 0.84) {$t$};
    \node[anchor=center] at (-3.459, 1.241) {$kt$};
    \node[anchor=center] at (3.994, 1.03) {$kt$};
    \node[anchor=center, font=\small, text=black!50] at (-1.354, 2.573) {$g$};
    \node[anchor=center, font=\small, text=black!50] at (2.71, 2.517) {$g$};
    \node[anchor=center, font=\small, text=black!50] at (-1.355, 1.592) {$h$};
    \node[anchor=center, font=\small, text=black!50] at (-3.04, 0.737) {$k$};
    \node[anchor=center, font=\small, text=black!50] at (4.157, 1.579) {$h$};
    \node[anchor=center, font=\small, text=black!50] at (4.255, 0.624) {$k$};
    \node[anchor=center, font=\small, text=black!50] at (-5.776, 1.219) {$ghk$};
    \node[anchor=center, font=\small, text=black!50] at (0.612, 1.837) {$ghk$};
    \node[anchor=center] at (-3.3, 2.269) {$ghkt$};
    \node[anchor=center] at (1.291, 2.211) {$ghkt$};
    \node[anchor=center] at (2.543, 1.671) {$hkt$};
    \node[anchor=center] at (-1.731, 2.059) {$hkt$};
    \node[anchor=center] at (-4.42, 3.358) {$gh,k(s)$};
    \node[anchor=center] at (-2.118, 3.44) {$g,h(kr)$};
    \node[anchor=center] at (-4.458, -0.118) {$gh,k(t)$};
    \node[anchor=center] at (-2.206, -0.125) {$g,h(kt)$};
    \node[anchor=center] at (2.584, 3.627) {${\omega}(g,h,k,s)$};
    \node[anchor=center] at (1.435, 3.006) {$g,hk(s)$};
    \node[anchor=center] at (3.668, 2.989) {$h,k(r)$};
    \node[anchor=center] at (1.358, 0.592) {$g,hk(t)$};
    \node[anchor=center] at (3.899, 0.175) {$(h,k)(t)$};
    \filldraw[thick, fill=RedOrange] (3.696, -0.635) rectangle (1.641, -0.127);
    \node[anchor=center] at (2.645, -0.393) {$\bar{\omega}(g,h,k,t)$};
\end{tikzpicture}
$$
Then it is clear that these associator satisfies \eqref{defomega} by vitue of the $4$-cocycle equation \eqref{4cocycleq}.
The proof that $O_g^\dagger = O_{g^{-1}}$ is shown in the next section. Even though the TN is different the calculation is the same.

\subsection{Unitary tensor network realization}\label{secTNU}
\begin{figure}[h!]
\centering
\includegraphics[width=0.8\textwidth]{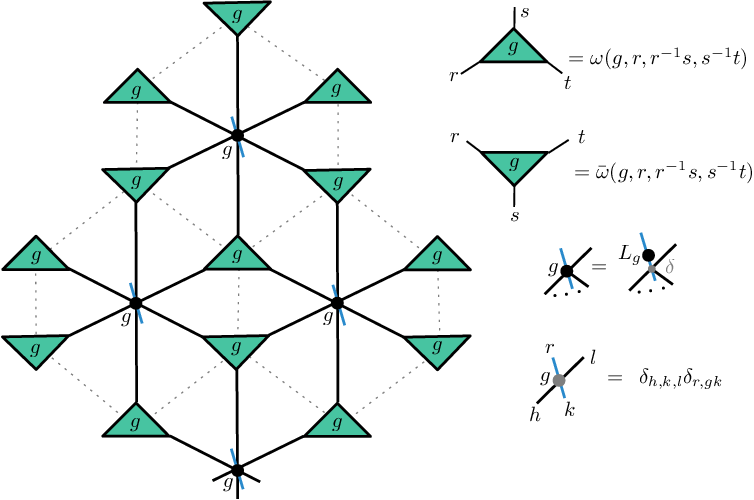}
\caption{The tensor network unitary representation of a finite group $G$ with a $4$-cocycle $\omega$. The different buiding tensors are shown: up and down tringle tensors carry the $4$-cocycle information, delta tensors depicted in gray and the $L_g$ in black. }
\label{fig:TNU}
\end{figure}

In this section we propose a modification of the previous construction that results in a unitary representation of the finite group $G$. To do so, we now place the physical indices on the vertices of a triangle lattice -- the dual of the hexagonal lattice -- valued also in $\mathbb{C}[G]$ such that the tensor network representation is shown in Fig.~\ref{fig:TNU}. The TN is composed of a net of delta tensors which are connected to the diagonal part carrying the 4-cocycle contribution, $\omega$, and the off-diagonal part corresponding to the left regular representation $L_g$. If $\omega=1$, then the unitary is just $L_g^{\otimes V}$.

{\bf Group structure.} Since the structure of this TN is not the same as the one studied in the previous section, we have to show that actually this is a representation of the group $G$ due to the fact that the local equations Eqs.~\eqref{prodTgh} and \eqref{orthoF} are not guaranteed to be satisfied now.

We will focus on a patch of the following form:

\begin{equation}\label{ExPatch}

\end{equation}

The fist step is to derive the form of the fusion MPO tensors. Those appear in pairs when multiplying the following configuration of tensors (the plaquette formed by two site tensors and two triangles, one pointing up and the other down):

\begin{equation}\label{pairedomega}
%
\end{equation}
The previous operation shows up three times in the patch shown in \eqref{ExPatch}. Gathering together the fusion MPO tensors that act on the middle triangle of \eqref{ExPatch}  we can conclude that they sum up to

\begin{equation}\label{omegared}
%
\end{equation}
which holds due to the $4$-cocycle equation:
$$
\omega(g,hr,r^{-1}s,s^{-1}t)\omega(h,r,r^{-1}s,s^{-1}t)\frac{{\omega}(g,h,r,r^{-1}t)}{{\omega}(g,h,r,r^{-1}s) {\omega}(g,h,s,s^{-1}t)} = \omega(gh,r,r^{-1}s,s^{-1}t) \ .
$$
We note that the unpaired tensors of 
Eq.~\eqref{omegared} coming from the decomposition of $\omega \cdot \bar{\omega}$ in Eq.~\eqref{pairedomega} correspond to the fusion MPO that appears at the boundary, since the multiplication $g\cdot h$ on the considered patch results in:
\begin{equation}
%
 \ ,
\end{equation}
where we have omitted the indices of $\omega$ and we have used the $4$-cocycle equation in the following form (that resembles a pulling through equation):

$$ %
$$
Hence the fusion MPO carries the $4$-cocycle index $\omega$ and it can be seen that satisfies the orthogonality relation Eq.~\eqref{orthoF}.\\

{\bf Unitarity:} So far we have shown that $U_g U_h = U_{gh}$. Using the normalized version of the $4$-cocycle where $\omega=1$ if any of its entries is the trivial group element $1$, it is easy to see that $U_1 = \id$, and then $U^{-1}_g = U_{g^{-1}}$. Let us now show that $U^{\dagger}_g = U_{g^{-1}} $. When we apply the dagger operation to $U_g$, the resulting tensors of the TN operator are:
$$
%
$$
Now, moving the left regular representations from the site tensors to the tensor represented by green triangle tensors, so that the site tensors are the ones from $U_{g^{-1}}$, we can see that the green triangle tensors transform as follows:
$$
%
 \ ,
$$
where we have defined the matrices $(w_g)_{r,s}=\omega(g,g^{-1},r,r^{-1}s)$ and the equality comes from the $4$-cocycle equation \eqref{4cocycleq} with the choice $(g,h, k, l, m)= (g, g^{-1},r, r^{-1}s, s^{-1}t)$ that reads:
$$ \bar{\omega}(g,g^{-1}r,  r^{-1}s,s^{-1}t) = {\omega}(g^{-1},r, r^{-1}s, s^{-1}t) \cdot \omega(g,g^{-1},r,r^{-1}t) \bar{\omega}(g,g^{-1},s,s^{-1}t) \bar{\omega}(g,g^{-1},r,r^{-1}s) \ . $$
The matrices $w_g$ are gauge transformations that cancel pairwise between site tensors:
$$
%
$$
Since the local tensors of the dagger of $U_g$ are related by a gauge transformation with the ones of $U_{g^{-1}}$, then $U^{\dagger}_g = U_{g^{-1}}$, i.e. $U_g$ is unitary.\\

{\bf Example:} Let us consider the  group $G=\mathbb{Z}_2\times \mathbb{Z}_2$. The local Hilbert space of the TNU is then two qubits that we denote by $A$ and $B$. There are four classes of 4-cocycles \cite{WangWencocycles}, since $H^4(\mathbb{Z}_2\times \mathbb{Z}_2,U(1))= \mathbb{Z}_2\times \mathbb{Z}_2$, that we index by $p^0,p^1=0,1$. The explicit formula for these four $4$-cocycles is
$$\omega(g,h,k,l) = (-1)^{p^0g_0h_1k_1l_1+p^1g_1h_0k_0l_0} \ ,$$
where we have denoted $g=(g_0,g_1)\in\mathbb{Z}_2\times \mathbb{Z}_2$. Labeling the group elements $(1,0)=a$ and $(0,1)=b$, the TNU of $p^0=p^1=1$ is composed of the following gates: 
$$
\begin{tikzpicture}[scale=1]
    \filldraw[very thick, fill=SeaGreen] (-1.292, 3.392) -- (-1.855, 2.828) -- (-0.731, 2.828) -- cycle;
    \node[anchor=center] at (-1.311, 3.035) {$a$};
    \node[anchor=center] at (-1.111, 3.941) {$|0\rangle_B$};
    \node[anchor=center] at (1.158, 2.883) {$= -1 $};
    \filldraw[shift={(4.349, 2.622)}, rotate=179.93, very thick, fill=SeaGreen] (0, 0) -- (-0.564, -0.564) -- (0.56, -0.564) -- cycle;
    \node[anchor=center] at (4.367, 2.962) {$a$};
    \draw[thick] (-1.281, 3.4) -- (-1.276, 3.754);
    \draw[thick] (-1.846, 2.842) -- (-2.183, 2.624);
    \draw[thick] (-0.739, 2.838) -- (-0.471, 2.633);
    \draw[thick] (4.349, 2.262) -- (4.354, 2.616);
    \draw[thick] (3.545, 3.375) -- (3.813, 3.171);
    \draw[thick] (5.27, 3.396) -- (4.933, 3.178);
    \node[anchor=center] at (-0.082, 2.468) {$|1\rangle_B$};
    \node[anchor=center] at (-2.496, 2.426) {$|1\rangle_B$};
    \node[anchor=center] at (3.192, 3.613) {$|1\rangle_B$};
    \node[anchor=center] at (5.695, 3.549) {$|1\rangle_B$};
    \node[anchor=center] at (4.368, 2.022) {$|0\rangle_B$};
    \node[anchor=center] at (6.288, 2.862) {$= -1 $};
    \filldraw[very thick, fill=SeaGreen] (-1.235, 0.891) -- (-1.799, 0.327) -- (-0.675, 0.327) -- cycle;
    \node[anchor=center] at (-1.255, 0.533) {$b$};
    \node[anchor=center] at (-1.055, 1.439) {$|0\rangle_A$};
    \node[anchor=center] at (1.214, 0.381) {$= -1 $};
    \filldraw[shift={(4.405, 0.121)}, rotate=179.93, very thick, fill=SeaGreen] (0, 0) -- (-0.564, -0.564) -- (0.56, -0.564) -- cycle;
    \node[anchor=center] at (4.423, 0.461) {$b$};
    \draw[thick] (-1.225, 0.898) -- (-1.22, 1.252);
    \draw[thick] (-1.79, 0.34) -- (-2.127, 0.123);
    \draw[thick] (-0.682, 0.336) -- (-0.415, 0.132);
    \draw[thick] (4.405, -0.239) -- (4.41, 0.115);
    \draw[thick] (3.601, 0.873) -- (3.869, 0.669);
    \draw[thick] (5.327, 0.894) -- (4.989, 0.677);
    \node[anchor=center] at (-0.026, -0.033) {$|1\rangle_A$};
    \node[anchor=center] at (-2.44, -0.076) {$|1\rangle_A$};
    \node[anchor=center] at (3.248, 1.111) {$|1\rangle_A$};
    \node[anchor=center] at (5.751, 1.048) {$|1\rangle_A$};
    \node[anchor=center] at (4.424, -0.479) {$|0\rangle_A$};
    \node[anchor=center] at (6.345, 0.36) {$= -1 $};
    \draw[RoyalBlue!80, very thick] (-6.143, 3.349) -- (-5.943, 2.667) -- (-5.943, 2.667);
    \node[disk large] at (-6.039, 2.963) {};
    \node[anchor=center] at (-4.909, 2.966) {$= X_A\otimes \mathbb{I}_B$};
    \node[anchor=center] at (-6.272, 2.99) {$a$};
    \draw[RoyalBlue!80, very thick] (-6.067, 0.671) -- (-5.867, -0.011) -- (-5.867, -0.011);
    \node[disk large] at (-5.963, 0.285) {};
    \node[anchor=center] at (-4.832, 0.288) {$=  \mathbb{I}_A \otimes X_B$};
    \node[anchor=center] at (-6.196, 0.312) {$b$};
\end{tikzpicture} \ ,
$$
where the have indicated the only non-zero contributions of the phase-gates $\omega$.

\subsection{Another TN realization: between the triple line and the unitary}

Here we show how we constructed the unitary operators from the triple line realization. From the structure of the TN of the triple line Eq.~\eqref{3Lview} we can think of another representation with less redundancy. To do so we first split each edge degree of freedom then, we correlate all the degrees of freedom in the same hexagonal plaquette. The resulting lattice structure is formed by a 6-party maximally entangled state in each plaquette:
$$
\begin{tikzpicture}[scale=1]
    \draw[thick] (5.655, -0.292) -- (5.09, -0.856) -- (5.09, -1.421) -- (5.655, -1.985) -- (6.219, -1.421) -- (6.219, -0.856) -- (5.655, -0.292);
    \draw[thick] (4.808, -0.856) -- (4.808, -1.421) -- (4.244, -1.985) -- (3.679, -1.421) -- (3.679, -0.856) -- (4.244, -0.292) -- (4.808, -0.856);
    \draw[thick] (4.387, 0.131) -- (4.387, -0.152) -- (4.951, -0.716) -- (5.515, -0.152) -- (5.515, 0.131);
    \draw[thick] (4.385, -2.408) -- (4.385, -2.126) -- (4.95, -1.562) -- (5.514, -2.126) -- (5.514, -2.408);
    \draw[thick] (4.103, 0.123) -- (4.103, -0.159) -- (3.538, -0.724) -- (3.256, -0.442);
    \draw[thick] (5.803, 0.141) -- (5.803, -0.141) -- (6.367, -0.705) -- (6.649, -0.423);
    \draw[thick] (5.793, -2.403) -- (5.793, -2.121) -- (6.357, -1.557) -- (6.64, -1.839);
    \draw[thick] (4.077, -2.384) -- (4.077, -2.102) -- (3.513, -1.537) -- (3.231, -1.82);
    \draw[thick] (3.104, -0.564) -- (3.387, -0.847) -- (3.387, -1.411) -- (3.104, -1.693);
    \draw[thick] (6.773, -0.564) -- (6.491, -0.847) -- (6.491, -1.411) -- (6.773, -1.693);
    \node[disk normal] at (3.679, -1.122) {};
    \node[disk normal] at (3.387, -1.122) {};
    \node[disk normal] at (3.933, -0.603) {};
    \node[disk normal] at (3.809, -0.453) {};
    \node[disk normal] at (4.646, -0.411) {};
    \node[disk normal] at (4.526, -0.574) {};
    \node[disk normal] at (4.808, -1.136) {};
    \node[disk normal] at (5.09, -1.121) {};
    \node[disk normal] at (5.358, -0.588) {};
    \node[disk normal] at (5.202, -0.465) {};
    \node[disk normal] at (5.923, -0.56) {};
    \node[disk normal] at (6.058, -0.397) {};
    \node[disk normal] at (6.491, -1.127) {};
    \node[disk normal] at (6.219, -1.127) {};
    \node[disk normal] at (6.064, -1.85) {};
    \node[disk normal] at (5.955, -1.685) {};
    \node[disk normal] at (5.372, -1.703) {};
    \node[disk normal] at (5.24, -1.852) {};
    \node[disk normal] at (4.67, -1.842) {};
    \node[disk normal] at (4.525, -1.704) {};
    \node[disk normal] at (3.962, -1.704) {};
    \node[disk normal] at (3.823, -1.847) {};
\end{tikzpicture}
$$
Since the vertices of  the hexagonal lattice of this TN still carry 3 indices, one for every plaquette, we can evaluate these with $\omega$ as in the triple line but now the off-diagonal part is placed at every edge:
$$
\begin{tikzpicture}[scale=1.5]
    \draw[thick] (5.655, -0.292) -- (5.09, -0.856) -- (5.09, -1.421) -- (5.655, -1.985) -- (6.219, -1.421) -- (6.219, -0.856) -- (5.655, -0.292);
    \draw[thick] (4.808, -0.856) -- (4.808, -1.421) -- (4.244, -1.985) -- (3.679, -1.421) -- (3.679, -0.856) -- (4.244, -0.292) -- (4.808, -0.856);
    \draw[thick] (3.104, -0.564) -- (3.387, -0.847) -- (3.387, -1.411) -- (3.104, -1.693);
    \filldraw[draw=RoyalBlue!80, fill=SeaGreen] (5.219, -1.721) -- (5.272, -2.018);
    \filldraw[draw=RoyalBlue!80, fill=SeaGreen] (5.353, -1.553) -- (5.406, -1.85);
    \filldraw[draw=RoyalBlue!80, fill=SeaGreen] (6.03, -1.71) -- (6.083, -2.007);
    \filldraw[draw=RoyalBlue!80, fill=SeaGreen] (5.931, -1.521) -- (5.984, -1.818);
    \filldraw[draw=RoyalBlue!80, fill=SeaGreen] (6.195, -0.984) -- (6.248, -1.282);
    \filldraw[draw=RoyalBlue!80, fill=SeaGreen] (6.472, -0.974) -- (6.525, -1.271);
    \filldraw[draw=RoyalBlue!80, fill=SeaGreen] (6.031, -0.239) -- (6.084, -0.536);
    \filldraw[draw=RoyalBlue!80, fill=SeaGreen] (5.898, -0.41) -- (5.951, -0.708);
    \filldraw[draw=RoyalBlue!80, fill=SeaGreen] (5.332, -0.451) -- (5.385, -0.748);
    \filldraw[draw=RoyalBlue!80, fill=SeaGreen] (5.175, -0.322) -- (5.227, -0.62);
    \filldraw[draw=RoyalBlue!80, fill=SeaGreen] (4.622, -0.27) -- (4.674, -0.567);
    \filldraw[draw=RoyalBlue!80, fill=SeaGreen] (4.504, -0.447) -- (4.557, -0.744);
    \filldraw[draw=RoyalBlue!80, fill=SeaGreen] (5.06, -0.976) -- (5.112, -1.274);
    \filldraw[draw=RoyalBlue!80, fill=SeaGreen] (4.787, -0.992) -- (4.839, -1.289);
    \filldraw[draw=RoyalBlue!80, fill=SeaGreen] (4.642, -1.706) -- (4.695, -2.003);
    \filldraw[draw=RoyalBlue!80, fill=SeaGreen] (4.501, -1.545) -- (4.554, -1.842);
    \filldraw[draw=RoyalBlue!80, fill=SeaGreen] (3.803, -1.71) -- (3.855, -2.007);
    \filldraw[draw=RoyalBlue!80, fill=SeaGreen] (3.938, -1.557) -- (3.991, -1.854);
    \filldraw[draw=RoyalBlue!80, fill=SeaGreen] (3.655, -0.998) -- (3.707, -1.295);
    \filldraw[draw=RoyalBlue!80, fill=SeaGreen] (3.36, -0.977) -- (3.413, -1.275);
    \filldraw[draw=RoyalBlue!80, fill=SeaGreen] (3.789, -0.303) -- (3.842, -0.601);
    \filldraw[draw=RoyalBlue!80, fill=SeaGreen] (3.91, -0.459) -- (3.962, -0.756);
    \draw[thick] (4.387, 0.131) -- (4.387, -0.152) -- (4.951, -0.716) -- (5.515, -0.152) -- (5.515, 0.131);
    \draw[thick] (4.385, -2.408) -- (4.385, -2.126) -- (4.95, -1.562) -- (5.514, -2.126) -- (5.514, -2.408);
    \draw[thick] (4.103, 0.123) -- (4.103, -0.159) -- (3.538, -0.724) -- (3.256, -0.442);
    \draw[thick] (5.803, 0.141) -- (5.803, -0.141) -- (6.367, -0.705) -- (6.649, -0.423);
    \draw[thick] (5.793, -2.403) -- (5.793, -2.121) -- (6.357, -1.557) -- (6.64, -1.839);
    \draw[thick] (4.077, -2.384) -- (4.077, -2.102) -- (3.513, -1.537) -- (3.231, -1.82);
    \draw[thick] (6.773, -0.564) -- (6.491, -0.847) -- (6.491, -1.411) -- (6.773, -1.693);
    \node[disk small] at (3.679, -1.122) {};
    \node[disk small] at (3.387, -1.122) {};
    \node[disk small] at (3.933, -0.603) {};
    \node[disk small] at (3.809, -0.453) {};
    \node[disk small] at (4.646, -0.411) {};
    \node[disk small] at (4.526, -0.574) {};
    \node[disk small] at (4.808, -1.136) {};
    \node[disk small] at (5.09, -1.121) {};
    \node[disk small] at (5.358, -0.588) {};
    \node[disk small] at (5.202, -0.465) {};
    \node[disk small] at (5.923, -0.56) {};
    \node[disk small] at (6.058, -0.397) {};
    \node[disk small] at (6.491, -1.127) {};
    \node[disk small] at (6.219, -1.127) {};
    \node[disk small] at (6.064, -1.85) {};
    \node[disk small] at (5.955, -1.685) {};
    \node[disk small] at (5.372, -1.703) {};
    \node[disk small] at (5.24, -1.852) {};
    \node[disk small] at (4.67, -1.842) {};
    \node[disk small] at (4.525, -1.704) {};
    \node[disk small] at (3.962, -1.704) {};
    \node[disk small] at (3.823, -1.847) {};
    \filldraw[fill=SeaGreen] (3.539, -0.736) -- (3.398, -0.841) -- (3.679, -0.85) -- cycle;
    \filldraw[fill=SeaGreen] (3.398, -1.413) -- (3.514, -1.525) -- (3.669, -1.415) -- cycle;
    \filldraw[fill=SeaGreen] (4.812, -1.422) -- (5.084, -1.421) -- (4.953, -1.558) -- cycle;
    \filldraw[fill=SeaGreen] (4.809, -0.859) -- (4.955, -0.712) -- (5.098, -0.859) -- cycle;
    \filldraw[fill=SeaGreen] (5.514, -0.145) -- (5.66, -0.304) -- (5.81, -0.143) -- cycle;
    \filldraw[fill=SeaGreen] (6.37, -0.706) -- (6.217, -0.852) -- (6.502, -0.852) -- cycle;
    \filldraw[fill=SeaGreen] (6.221, -1.42) -- (6.497, -1.419) -- (6.36, -1.569) -- (6.36, -1.569) -- cycle;
    \filldraw[fill=SeaGreen] (5.655, -1.985) -- (5.515, -2.127) -- (5.803, -2.124) -- cycle;
    \filldraw[fill=SeaGreen] (4.243, -1.979) -- (4.081, -2.102) -- (4.397, -2.116) -- cycle;
    \filldraw[fill=SeaGreen] (4.103, -0.153) -- (4.387, -0.152) -- (4.241, -0.293) -- cycle;
\end{tikzpicture}
$$
where:
$$
\begin{tikzpicture}[scale=1]
    \filldraw[very thick, fill=SeaGreen] (4.219, 4.738) -- (3.655, 4.175) -- (4.779, 4.175) -- cycle;
    \draw[very thick] (3.494, 4.719) -- (4, 4.27) -- (3.999, 3.663);
    \draw[very thick] (3.729, 4.926) -- (4.218, 4.487) -- (4.706, 4.914);
    \draw[very thick] (4.916, 4.752) -- (4.391, 4.301) -- (4.388, 3.687);
    \node[anchor=center] at (4.199, 4.338) {$g$};
    \node[anchor=center] at (3.4, 4.566) {$h$};
    \node[anchor=center] at (3.844, 3.709) {$h$};
    \node[anchor=center] at (4.556, 3.707) {$l$};
    \node[anchor=center] at (4.987, 4.562) {$l$};
    \node[anchor=center] at (3.889, 5.024) {$k$};
    \node[anchor=center] at (4.505, 5.028) {$k$};
    \node[anchor=center] at (6.668, 4.229) {$= \omega(g,h,h^{-1}k,k^{-1}l)$};
\end{tikzpicture}
$$
Finally, since the action of the operators on the same plaquette is correlated, we can view it as a single action on the plaquette, associating its Hilbert space in the face, and then we arrive to the unitary representation of Fig.~\ref{fig:TNU}.

\section{(2+1)D anomalous group symmetric phases}

In this section we promote the TN operators described in Sec.~\ref{TNOG} to symmetries of two-dimensional tensor networks, i.e. projected entangled pairs state (PEPS) \cite{Verstraete04}. We think of these PEPS as the ground states of gapped Hamiltonians so that our characterization stands for classifying the quantum phases of these symmetries. 

Let us consider a set of PEPS, $\ket{\psi_x}$ where $x\in \mathcal{X}$ is the label set, that we take as the basis of the ground space so that the degeneracy is $|\mathcal{X}|$. The PEPS are placed on a hexagonal lattice, as the symmetry operators, constructed with tensors $A_x$, for all $x\in \mathcal{X}$:

\begin{equation}
    \ket{\psi_x} = \ 
\begin{tikzpicture}[scale=1,baseline=-1cm]
    \draw[shift={(5.434, -1.023)}, xscale=0.588, yscale=0.617, RoyalBlue!80, very thick] (0, 0) -- (0.269, -0.732);
    \draw[shift={(4.59, -1.6)}, xscale=0.522, yscale=0.58, RoyalBlue!80, very thick] (0, 0) -- (0.269, -0.732);
    \draw[shift={(4.575, 0.663)}, xscale=0.593, yscale=0.576, RoyalBlue!80, very thick] (0, 0) -- (0.269, -0.732);
    \draw[shift={(5.423, 0.114)}, xscale=0.557, yscale=0.596, RoyalBlue!80, very thick] (0, 0) -- (0.269, -0.732);
    \draw[shift={(3.729, -1.021)}, xscale=0.548, yscale=0.6, RoyalBlue!80, very thick] (0, 0) -- (0.269, -0.732);
    \draw[shift={(3.753, 0.074)}, xscale=0.466, yscale=0.493, RoyalBlue!80, very thick] (0, 0) -- (0.269, -0.732);
    \draw[very thick, densely dashed] (3.878, -0.286) -- (3.878, -1.415) -- (4.725, -1.98) -- (5.571, -1.415) -- (5.571, -0.286) -- (4.725, 0.278) -- cycle;
    \draw[very thick, densely dashed] (4.725, 0.278) -- (4.725, 0.843);
    \draw[very thick, densely dashed] (3.878, -0.286) -- (3.031, 0.278);
    \draw[very thick, densely dashed] (5.571, -0.286) -- (6.418, 0.278);
    \draw[very thick, densely dashed] (5.571, -1.415) -- (6.418, -1.98);
    \draw[very thick, densely dashed] (3.878, -1.415) -- (3.031, -1.98);
    \draw[very thick, densely dashed] (4.725, -1.98) -- (4.725, -2.544);
    \node[disk normal] at (3.878, -0.286) {};
    \node[disk normal] at (4.725, 0.278) {};
    \node[disk normal] at (5.571, -0.286) {};
    \node[disk normal] at (3.878, -1.415) {};
    \node[disk normal] at (4.725, -1.98) {};
    \node[disk normal] at (5.571, -1.415) {};
    \node[anchor=center] at (4.254, -0.463) {$A_x$};
\end{tikzpicture}
\end{equation}

We now consider the TNOs of Eq.~\eqref{Og} as symmetries of our ground space spanned by $\{ \ket{\psi_x}, x\in \mathcal{X}\} $. In general the action will be given given by a permutation:
$$ O_g \ket{\psi_x} = \ket{\psi_y}\ ,$$
where we define a group action on the ground state basis set: $y\equiv g\cdot x $. 
The main assumption here is how the local symmetry tensor $T_g$ acts on $A_x$: we assume that there exists an {\it action MPO} that realizes that permutation locally:

$$\begin{tikzpicture}[scale=1]
    \draw[very thick] (2.487, 0.043) -- (2.869, 0.514) -- (2.869, 0.514);
    \draw[very thick, densely dashed] (2.499, -0.038) -- (3.171, -0.256);
    \draw[very thick, densely dashed] (-0.005, -0.157) -- (-0.652, 0.04);
    \draw[very thick] (0.021, -0.137) -- (-0.203, 0.639);
    \draw[very thick] (0.412, -2.089) -- (-0.251, -2.164);
    \draw[shift={(0.417, -2.109)}, xscale=1.273, yscale=1.307, very thick, densely dashed] (0, 0) -- (-0.273, -0.472) -- (-0.273, -0.472);
    \draw[very thick] (4.43, 0.984) -- (5.559, 0.137) -- (6.688, 0.984) -- (6.688, 0.984);
    \draw[very thick] (5.559, 0.137) -- (4.995, -0.991) -- (4.995, -0.991);
    \draw[RoyalBlue!80, very thick] (5.559, 0.702) -- (5.559, -0.427) -- (5.559, -0.709);
    \node[disk normal] at (5.559, 0.137) {};
    \draw[very thick, densely dashed] (4.43, -0.427) -- (5.559, -1.274) -- (6.688, -0.427) -- (6.688, -0.427);
    \draw[very thick, densely dashed] (5.559, -1.274) -- (4.995, -2.403) -- (4.995, -2.403);
    \draw[shift={(5.559, -0.709)}, yscale=0.413, RoyalBlue!80, very thick] (0, 0) -- (0, -1.129) -- (0, -1.411);
    \node[disk normal] at (5.559, -1.274) {};
    \draw[shift={(1.118, -0.702)}, scale=1.25, very thick, densely dashed] (0, 0) -- (-0.564, -1.129) -- (-0.564, -1.129);
    \draw[shift={(1.118, -0.138)}, yscale=0.418, RoyalBlue!80, very thick] (0, 0) -- (0, -1.129) -- (0, -1.411);
    \node[disk normal] at (1.118, -0.702) {};
    \node[anchor=center, font=\LARGE] at (3.531, -0.759) {$=$};
    \draw[shift={(2.47, 0.002)}, xscale=1.25, Green!80, very thick, densely dashed] (0, 0) .. controls (0.047, -0.706) and (-1.082, -2.117) .. (-1.693, -2.117) .. controls (-2.305, -2.117) and (-2.399, -0.706) .. (-1.834, 0) .. controls (-1.27, 0.706) and (-0.047, 0.706) .. cycle;
    \draw[very thick, densely dashed] (1.124, -0.692) -- (-0.005, -0.128);
    \draw[very thick, densely dashed] (1.118, -0.702) -- (2.47, 0.004) -- (2.47, 0.004);
    \node[disk normal] at (2.47, 0.004) {};
    \node[anchor=center] at (5.915, -0.138) {$T_g
$};
    \node[anchor=center] at (5.915, -1.549) {$A_x$};
    \node[anchor=center] at (1.511, -0.92) {$A_{gx}$};
    \node[anchor=center] at (0.284, -0.587) {$W_{g,x}$};
    \node[anchor=center] at (1.04, 0.765) {$g,x
$};
    \node[fsquare large, fill=white] at (0.036, -0.161) {};
    \node[fsquare large, fill=white] at (2.466, 0.004) {};
    \node[fsquare large, fill=white] at (0.411, -2.101) {};
\end{tikzpicture}
$$
We also assume a local orthogonality of these action MPO tensors:
$$
\begin{tikzpicture}[scale=0.8]
    \draw[Green!80, very thick, densely dashed] (-4.282, -0.67) -- (-4.282, -3.21);
    \draw[Green!80, very thick, densely dashed] (-3.066, -0.687) -- (-3.066, -3.227);
    \draw[very thick, densely dashed] (-4.254, -1.935) -- (-4.818, -1.935) -- (-5.1, -1.935);
    \draw[very thick, densely dashed] (-2.233, -1.933) -- (-2.798, -1.933) -- (-3.08, -1.933);
    \node[anchor=center, font=\LARGE] at (-1.842, -1.92) {$=$};
    \draw[shift={(1.595, -1.895)}, xscale=2.088, yscale=0.164, very thick, densely dashed] (0, 0) -- (-0.564, 0) -- (-0.847, 0);
    \draw[Green!80, very thick, densely dashed] (0.144, -0.903) .. controls (0.131, -1.729) and (1.238, -1.748) .. (1.258, -0.905);
    \draw[shift={(1.305, -2.904)}, rotate=-179.913, Green!80, very thick, densely dashed] (0, 0) .. controls (-0.013, -0.827) and (1.095, -0.845) .. (1.115, -0.003);
    \draw[very thick] (-4.251, -1.934) -- (-3.665, -1.137) -- (-3.085, -1.912);
    \draw[very thick, densely dashed] (-4.235, -1.96) -- (-3.672, -2.824) -- (-3.074, -1.958);
    \node[fsquare large, fill=white] at (-4.267, -1.938) {};
    \node[fsquare large, fill=white] at (-3.065, -1.939) {};
    \node[anchor=center] at (-4.551, -0.879) {$g,x$};
    \node[anchor=center] at (-2.76, -0.928) {$h,y$};
    \node[anchor=center] at (-0.843, -1.908) {$\delta_{g,h}\delta_{x,y}$};
    \node[anchor=center] at (0.738, -1.252) {$g,x$};
    \node[anchor=center] at (-4.808, -1.751) {$gx$};
    \node[anchor=center] at (-2.526, -1.739) {$hy$};
    \node[anchor=center] at (1.942, -1.904) {$gx$};
    \node[anchor=center] at (-3.873, -1.72) {$g$};
    \node[anchor=center] at (-3.844, -2.249) {$x$};
    \node[anchor=center] at (-3.409, -1.696) {$h
$};
    \node[anchor=center] at (-3.454, -2.236) {$y$};
    \node[anchor=center] at (0.772, -2.599) {$g,x$};
\end{tikzpicture} \ ,
$$
which guarantees that the symmetry action on a PEPS patch behaves properly in the bulk and just the boundary is left with the action MPO.

Associativity, and the invertibility of the PEPS tensors, implies the equality between two action MPOs and a combination of a fusion and an action MPO. That equation is satisfied by assuming the existence of some mixed associator tensors depicted in green:
$$
\begin{tikzpicture}[scale=1,baseline=1cm]
    \draw[shift={(7.334, -0.322)}, xscale=6.248, yscale=0.15, RedOrange, very thick] (0, 0) -- (0, -3.387);
    \draw[shift={(7.31, 2.781)}, xscale=6.248, yscale=0.15, RedOrange, very thick] (0, 0) -- (0, -3.387);
    \draw[shift={(7.312, 2.026)}, xscale=-0.086, yscale=0.704, Green!80, very thick, densely dashed] (0, 0) -- (0, -3.387);
    \draw[shift={(3.093, 2.765)}, xscale=1.805, yscale=1.067, RedOrange, very thick] (0, 0) -- (0, -3.387);
    \draw[shift={(6.485, 2.764)}, xscale=1.805, yscale=1.067, Green!80, very thick, densely dashed] (0, 0) -- (0, -3.387);
    \draw[shift={(2.284, 2.735)}, xscale=1.805, yscale=1.067, Green!80, very thick, densely dashed] (0, 0) -- (0, -3.387);
    \node[anchor=center, font=\LARGE] at (4.789, 0.753) {$=$};
    \filldraw[shift={(6.118, 2.402)}, xscale=1.009, yscale=1.19, fill=Green!80] (0, 0) rectangle (1.401, -0.36);
    \filldraw[shift={(6.153, -0.111)}, yscale=1.145, fill=Green!80] (0, 0) rectangle (1.401, -0.36);
    \node[anchor=center] at (6.8, 2.156) {$G_{g,h,x}$};
    \node[anchor=center] at (6.8, -0.3) {$\tilde{G}_{g,h,x}$};
    \node[anchor=center] at (3.465, 2.238) {$g,h$};
    \node[anchor=center] at (1.806, 2.284) {$gh,x$};
    \node[anchor=center] at (5.918, 1.673) {$g,hx$};
    \node[anchor=center] at (7.691, 1.634) {$h,x$};
    \draw[very thick, densely dashed] (6.492, 1.132) -- (5.928, 1.132) -- (5.646, 1.132);
    \draw[very thick, densely dashed] (6.492, 1.132) -- (7.339, 0.568);
    \draw[very thick] (6.492, 1.132) -- (7.057, 1.697) -- (7.057, 1.697);
    \draw[very thick] (7.312, 0.595) -- (7.877, 1.16) -- (7.877, 1.16);
    \draw[very thick, densely dashed] (7.323, 0.579) -- (7.889, 0.303) -- (7.889, 0.303);
    \node[fsquare large, fill=white] at (6.495, 1.117) {};
    \node[fsquare large, fill=white] at (7.336, 0.6) {};
    \draw[very thick] (3.668, 1.702) -- (3.103, 1.137) -- (3.668, 0.573);
    \draw[very thick, densely dashed] (2.256, 0.573) -- (1.692, 0.573) -- (1.41, 0.573);
    \node[disk normal] at (3.103, 1.137) {};
    \node[anchor=center] at (3.95, 1.702) {$g$};
    \node[anchor=center] at (3.95, 0.573) {$h$};
    \node[anchor=center] at (2.972, 0.036) {$x$};
    \draw[very thick] (3.088, 1.123) -- (2.293, 0.589);
    \draw[very thick, densely dashed] (2.275, 0.556) -- (2.841, 0.28) -- (2.841, 0.28);
    \node[fsquare large, fill=white] at (2.263, 0.576) {};
    \node[anchor=center] at (7.699, 2.803) {$g,h$};
    \node[anchor=center] at (6.039, 2.566) {$gh,x$};
\end{tikzpicture}
\  \Rightarrow \ 
\begin{tikzpicture}[scale=1, baseline=2cm]
    \draw[shift={(-5.479, 4.924)}, yscale=1.333, Green!80, very thick, densely dashed] (0, 0) -- (0, -3.387) -- (0, -3.387);
    \draw[shift={(-4.632, 4.924)}, yscale=1.333, RedOrange, very thick] (0, 0) -- (0, -3.387);
    \draw[very thick] (-4.068, 4.641) -- (-4.632, 4.077) -- (-4.068, 3.512);
    \draw[very thick, densely dashed] (-5.479, 3.512) -- (-6.044, 3.512) -- (-6.326, 3.512);
    \node[disk normal] at (-4.632, 4.077) {};
    \draw[shift={(-1.528, 4.924)}, yscale=1.333, Green!80, very thick, densely dashed] (0, 0) -- (0, -3.387) -- (0, -3.387);
    \draw[shift={(-0.681, 4.924)}, yscale=1.333, Green!80, very thick, densely dashed] (0, 0) -- (0, -3.387);
    \draw[very thick, densely dashed] (-1.528, 4.077) -- (-2.092, 4.077) -- (-2.375, 4.077);
    \draw[very thick, densely dashed] (-1.528, 4.077) -- (-0.681, 3.512);
    \draw[very thick] (-1.528, 4.077) -- (-0.964, 4.641) -- (-0.964, 4.641);
    \node[anchor=center, font=\LARGE] at (-2.939, 2.666) {$=$};
    \draw[very thick] (-4.068, 2.384) -- (-4.632, 1.819) -- (-4.068, 1.255);
    \draw[very thick, densely dashed] (-5.479, 1.255) -- (-6.044, 1.255) -- (-6.326, 1.255);
    \node[disk normal] at (-4.632, 1.819) {};
    \node[anchor=center] at (-3.786, 4.641) {$g$};
    \node[anchor=center] at (-3.786, 3.512) {$h$};
    \node[anchor=center] at (-4.764, 2.975) {$x$};
    \node[anchor=center] at (-4.615, 0.296) {$g,h$};
    \node[anchor=center] at (-5.542, 0.26) {$gh,x$};
    \node[anchor=center] at (-0.642, 0.193) {$h,x$};
    \node[anchor=center] at (-1.612, 0.208) {$g,hx$};
    \draw[very thick] (-4.648, 4.062) -- (-5.442, 3.528);
    \draw[very thick, densely dashed] (-5.46, 3.495) -- (-4.894, 3.219) -- (-4.894, 3.219);
    \draw[very thick] (-4.678, 1.8) -- (-5.473, 1.266);
    \draw[very thick, densely dashed] (-5.465, 1.228) -- (-4.899, 0.952) -- (-4.899, 0.952);
    \node[fsquare large, fill=white] at (-5.473, 3.516) {};
    \node[fsquare large, fill=white] at (-5.482, 1.236) {};
    \draw[very thick] (-0.708, 3.54) -- (-0.144, 4.104) -- (-0.144, 4.104);
    \draw[very thick, densely dashed] (-0.697, 3.523) -- (-0.131, 3.247) -- (-0.131, 3.247);
    \node[fsquare large, fill=white] at (-1.526, 4.062) {};
    \node[fsquare large, fill=white] at (-0.685, 3.545) {};
    \draw[very thick, densely dashed] (-1.518, 1.623) -- (-2.083, 1.623) -- (-2.365, 1.623);
    \draw[very thick, densely dashed] (-1.518, 1.623) -- (-0.672, 1.058);
    \draw[very thick] (-1.518, 1.623) -- (-0.954, 2.187) -- (-0.954, 2.187);
    \draw[very thick] (-0.698, 1.085) -- (-0.134, 1.65) -- (-0.134, 1.65);
    \draw[very thick, densely dashed] (-0.688, 1.069) -- (-0.121, 0.793) -- (-0.121, 0.793);
    \node[fsquare large, fill=white] at (-1.516, 1.607) {};
    \node[fsquare large, fill=white] at (-0.675, 1.091) {};
\end{tikzpicture}
$$
These tensors are defined up to a phase factor:
$$G(g,h,x)\to \beta(g,h,x) G(g,h,x) \ , \ \tilde{G}(g,h,x)\to \bar{\beta}(g,h,x) \tilde{G}(g,h,x)  \ ,$$
where the product of them gives the identity on the corresponding spaces:
$$ \tilde{G}(g,h,x)\cdot {G}(g,h,x) = \id_{g,hx}\otimes \id_{h,x}\ .$$

The pentagon relation implies that the following combination of associator tensors is a constant factor labeled as $\Lambda$:

$$
\begin{tikzpicture}[scale=1]
    \draw[Green!80, very thick, densely dashed] (3.119, -0.29) -- (-4.432, -0.333);
    \draw[shift={(1.15, -1.172)}, xscale=0.071, yscale=-0.008, Green!80, very thick, densely dashed] (0, 0) -- (-7.551, -0.043);
    \draw[shift={(-3.935, -2.044)}, xscale=0.071, yscale=-0.008, Green!80, very thick, densely dashed] (0, 0) -- (-7.551, -0.043);
    \draw[shift={(1.968, -1.26)}, xscale=1.565, yscale=0.025, RedOrange, very thick] (0, 0) -- (0.364, 0.378);
    \node[anchor=center] at (-0.039, -1.825) {$g,h$};
    \node[anchor=center] at (-0.232, -0.998) {$gh,k$};
    \node[anchor=center] at (-1.784, -0.993) {$g,hk$};
    \node[anchor=center] at (-2.308, -1.825) {$h,k$};
    \filldraw[shift={(-2.7, -1.437)}, rotate=89.772, fill=Green!80] (0, 0) rectangle (1.401, -0.36);
    \filldraw[shift={(-1.268, -2.338)}, rotate=89.772, fill=RedOrange] (0, 0) rectangle (1.401, -0.36);
    \filldraw[shift={(0.302, -1.43)}, rotate=89.772, fill=Green!80] (0, 0) rectangle (1.401, -0.36);
    \node[anchor=center] at (-4.744, -2.022) {$k,x$};
    \node[anchor=center] at (-4.834, -1.21) {$h,kx$};
    \node[anchor=center] at (-4.92, -0.339) {$g,hkx$};
    \node[anchor=center] at (-3.203, -1.034) {$hk,x$};
    \node[anchor=center] at (-1.083, -0.085) {$ghk,x$};
    \node[anchor=center] at (3.409, -2.027) {$k,x$};
    \node[anchor=center] at (1.616, -0.128) {$gh,kx$};
    \draw[Green!80, very thick, densely dashed] (1.161, -1.184) -- (2.027, -2.018) -- (3.162, -2.024);
    \draw[shift={(1.209, -2.024)}, xscale=2.1, yscale=2.038, RedOrange, very thick] (0, 0) -- (0.364, 0.378);
    \filldraw[shift={(2.483, -1.419)}, rotate=89.772, fill=Green!80] (0, 0) rectangle (1.401, -0.36);
    \node[anchor=center] at (3.528, -1.208) {$h,kx$};
    \node[anchor=center] at (3.63, -0.289) {$g,hkx$};
    \node[anchor=center] at (6.5, -1.069) {$= \Lambda_{g,h,k}^x \cdot \id_{k,x} \otimes \id_{h,kx}\otimes \id_{g,hkx}$};
    \filldraw[shift={(-4.023, -2.326)}, rotate=89.772, fill=Green!80] (0, 0) rectangle (1.401, -0.36);
    \draw[RedOrange, very thick] (0.296, -1.201) -- (-2.33, -1.2);
    \draw[shift={(-2.711, -1.203)}, xscale=0.231, yscale=0.005, Green!80, very thick, densely dashed] (0, 0) -- (-7.551, -0.043);
    \draw[RedOrange, very thick] (-3.644, -2.052) -- (1.214, -2.036);
    \draw[shift={(3.141, -1.235)}, xscale=0.071, yscale=-0.008, Green!80, very thick, densely dashed] (0, 0) -- (-7.551, -0.043);
\end{tikzpicture}
$$
This structure can be seen as the 3D projection of a 4-simplex valued as follows:
\begin{equation}\label{4simplexLambda}
\begin{tikzpicture}[scale=1]
    \draw[RedOrange, thick] (-4.297, 2.124) -- (0.216, 2.126);
    \draw[white, very thick] (-2.06, 2.267) -- (-2.06, 1.984);
    \draw[white, very thick] (-1.99, 2.26) -- (-1.99, 1.978);
    \draw[white, very thick] (-2.708, 2.261) -- (-2.81, 1.958);
    \draw[white, very thick] (-2.767, 2.282) -- (-2.869, 1.978);
    \node[anchor=center] at (-3.384, 3.679) {$g,hkx$};
    \node[anchor=center] at (-0.393, 3.314) {$g,hk$};
    \node[anchor=center] at (-1.35, 1.278) {$h,k$};
    \node[anchor=center] at (-0.367, 0.438) {$gh,k$};
    \node[anchor=center] at (-3.372, 2.208) {$g,h$};
    \node[anchor=center] at (-3.489, 0.133) {$gh,kx$};
    \node[anchor=center] at (-2.033, 4.71) {$G(g,hk,x)$};
    \node[anchor=center] at (-0.992, -1.222) {$\tilde{G}(gh,k,x)$};
    \node[anchor=center] at (0.854, 2.451) {$R(g,h,k)$};
    \node[anchor=center] at (-5.033, 2.462) {$\tilde{G}(g,h,kx)$};
    \node[anchor=center] at (-4.219, 0.82) {$G(h,k,x)$};
    \node[anchor=center, font=\large] at (2.702, 2.074) {$= \Lambda^x_{g,h,k} \cdot \chi$};
    \draw[Green!80, thick] (-3.168, 0.995) -- (-4.297, 2.124) -- (-2.039, -1.263) -- (-3.168, 0.995);
    \draw[RedOrange, thick] (-2.011, 4.391) -- (0.219, 2.124) -- (-2.042, -1.259);
    \draw[Green!80, thick] (-3.168, 0.995) -- (-2.011, 4.391);
    \draw[Green!80, thick] (-4.297, 2.124) -- (-2.011, 4.391) -- (-2.042, -1.262);
    \draw[shift={(-2.09, 1.317)}, xscale=0.945, yscale=1.326, white, very thick] (0, 0) -- (0.155, 0.036) -- (0.155, 0.036);
    \draw[shift={(-2.106, 1.388)}, xscale=0.956, yscale=1.387, white, very thick] (0, 0) -- (0.155, 0.036) -- (0.155, 0.036);
    \draw[RedOrange, thick] (0.216, 2.126) -- (-3.168, 0.995);
    \draw[Green!80, thick] (-3.168, 0.995) -- (-4.297, 2.124);
    \node[anchor=center] at (-1.621, 2.744) {$ghk,x$};
    \node[anchor=center] at (-2.79, 2.786) {$hk,x$};
    \node[anchor=center] at (-3.436, 1.596) {$h,kx$};
    \node[anchor=center] at (-2.588, 0.397) {$k,x$};
    \node[fsquare large, fill=Green!80] at (-2.039, 4.382) {};
    \node[fsquare large, fill=Green!80] at (-2.039, -1.263) {};
    \node[fsquare large, fill=Green!80] at (-4.297, 2.124) {};
    \node[fsquare large, fill=BrickRed!90] at (0.219, 2.124) {};
    \node[fsquare large, fill=Green!80] at (-3.168, 0.995) {};
\end{tikzpicture}
\ ,
\end{equation}
where $\chi$ is the product of the dimensions of the virtual spaces ${(k,x)}$, ${(h,kx)}$ and  ${(g,hkx)}$.

The quantity $\Lambda$ is defined up to the following equivalence relation due to the freedom of the action associators: 
\begin{equation}\label{Lambdafreedom}
\Lambda_{g, h, k}^ {x }\to \frac{\beta(g,hk,x)\beta(h,k,x)}{\beta(gh,k,x)\beta(g,h,kx)}\Lambda_{g, h, k}^ {x } \ ;
\end{equation}
note that there is no freedom from the fusion associator $R$ since this is fixed when choosing the symmetry tensors. 

The associahedron relation implies the following mixed 4-cocycle condition for $\Lambda$ and $\omega$:
\begin{equation} \label{mixed4cocycle}
\Lambda_{g, h, k}^ {lx }  \Lambda_{g, h k, l}^ { x}  \Lambda_{h, k, l}^ { x}
= \Lambda_{g, h, k l}^ { x}  \Lambda_{g h, k, l}^ { x} \omega(g, h, k, l) \  .
\end{equation}

The different solutions to this equation, inequivalent $\Lambda$'s based on the freedom of \eqref{Lambdafreedom}, encode the different 2D quantum phases symmetric under a global TNO representation of $G$ with a $4$-cocycle index $\omega$. Let us comment on some implications of this equation.

Let us suppose that there is a unique ground state, $|\mathcal{X}|=1$, labeled by $x$. 
Then Eq.\eqref{mixed4cocycle} can be written as
$$\omega(g, h, k, l)=\frac{\Lambda_{g, h, k}^ {x }  \Lambda_{g, h k, l}^ { x}  \Lambda_{h, k, l}^ { x}}{ \Lambda_{g, h, k l}^ { x}  \Lambda_{g h, k, l}^ { x}} \ , $$
which implies that $\omega$ can be written as a 4-coboundary, i.e. it is trivial, $\omega\in [1] \in H^4(G,U(1))$. Therefore, the 4-cocycle $\omega$ corresponds to the anomaly of the tensor network symmetry: a non-trivial index prevents the Hamiltonian from having a unique ground state.

Let us denote by $H\subseteq G$ the unbroken symmetry group such that $H\cdot x =x$ for all $x \in \mathcal{X}$.  Then the ground state degeneracy is $|G/H|$. The elements in the set $\in G-H$ are the only ones encoding the permutation of the ground states. For the elements of the unbroken group we can see from Eq.\eqref{mixed4cocycle} that $\omega$ becomes trivial. Then, for a suitable gauge choice $\Lambda_{g, h, k}^ {x }$ is a $3$-cocycle of $H$ that classifies the SPT phase of the unbroken symmetry part. This concludes the characterization of phases invariant under anomalous symmetries.

We note that we can also provide representations of these PEPS by using the same approach as in the triple line representation of the symmetries: using \eqref{mixed4cocycle} as a guide to assign $\Lambda$ and $\omega$ values to the tensors.

With the same techniques that we developped in Ref.~\cite{Garre22_MPOSYM}, we could show that a differentiable change of the PEPS tensor, which is invariant under the same symmetry operators, will modify the action MPOs in such a way that the new action assocciators will define a $\Lambda$ that belongs to the same class as the previous one. This shows that the equivalence classes of $\Lambda$ are invariants of the phase. Moreover, we can construct a continuous and symmetric path that connects two PEPS that are symmetric under possibly different
representations of a group $G$, provide the two representation have the same $4$-cocycle index, and whose symmetry action is characterized by the same class of
$\Lambda$. Therefore two of such PEPS belong to the same quantum phase.

\section{Symmetry Protected Topological phases based on TN}

In Ref.\cite{Chen13} the classification and construction of SPT phases based on group cohomology was achieved. In this section we focus on the TN-based approach of $(3+1)$D SPT phases. Due to the similarities with its construction, we first review  the $(2+1)$D SPT case based on TN.
\subsection{The (2+1) SPT case based on TN}\label{2dSPTcase}

In this section we review the (2+1) SPT models based on group cohomology of \cite{Chen13}, focusing on their TN representation. The symmetries of these models are based on matrix product unitaries (MPUs), $U_g$ for all $g\in G$ with tensors $T_g$:
$$ U_g =
\begin{tikzpicture}[scale=1, baseline=3cm]
    \draw[RedOrange, thick] (5.08, 3.387) -- (8.184, 3.387);
    \draw[thick] (5.362, 2.822) -- (5.362, 3.951);
    \draw[thick] (5.927, 3.951) -- (5.927, 2.822);
    \draw[thick] (6.491, 3.951) -- (6.491, 2.822);
    \draw[thick] (7.056, 3.951) -- (7.056, 2.822);
    \draw[thick] (7.62, 3.951) -- (7.62, 2.822);
    \draw[RedOrange, thick] (5.08, 3.387) -- (4.798, 3.387) -- (4.798, 2.54) -- (8.184, 2.54) -- (8.184, 3.387);
    \node[disk normal] at (5.362, 3.387) {};
    \node[disk normal] at (5.927, 3.387) {};
    \node[disk normal] at (6.491, 3.387) {};
    \node[disk normal] at (7.056, 3.387) {};
    \node[disk normal] at (7.62, 3.387) {};
    \node[anchor=center] at (5.08, 3.104) {$T_g$};
\end{tikzpicture}
 \ .
$$
They are constructed a given $3$-cocycle $\omega$ that satisfies:
\begin{equation}\label{3cocyleeq}
\omega(g, h, k)\omega(g, hk, l)\omega(h, k, l) = \omega(gh, k, l) \omega(g, h, kl)
\end{equation}

The MPU tensors are the following:

$$
\begin{tikzpicture}[scale=1]
    \draw[RedOrange, thick] (-6.223, 4.328) -- (-4.811, 4.328);
    \draw[thick] (-5.658, 4.893) -- (-5.658, 3.764);
    \node[disk normal] at (-5.658, 4.61) {};
    \node[disk normal, Green!80] at (-5.094, 4.328) {};
    \node[anchor=center] at (-6, 4.586) {$L_g$};
    \node[anchor=center] at (-5.157, 4) {$\omega_g$};
    \node[anchor=center, font=\footnotesize] at (-6.4, 4.318) {$r$};
    \node[anchor=center, font=\footnotesize] at (-4.698, 4.313) {$s$};
    \node[anchor=center, font=\footnotesize] at (-5.665, 3.55) {$r$};
    \node[anchor=center, font=\footnotesize] at (-5.662, 4.983) {$gr$};
    \node[anchor=center] at (-3.336, 4.2) {$= \omega(g,r,r^{-1}s)$};
    \node[anchor=center] at (-7, 4.2) {$T_g = $};
\end{tikzpicture} \ .
$$
The operators satisfy the group relations $U_gU_h=U_{gh}$ via the following local decomposition:
\begin{equation}\label{tgthdecomp}
\begin{tikzpicture}[scale=1]
    \draw[RedOrange, thick] (3.27, 1.795) -- (3.835, 1.795) -- (4.399, 2.077);
    \draw[RedOrange, thick] (-0.564, 1.693) -- (0, 1.693);
    \draw[thick] (-5.927, 2.54) -- (-5.927, 1.129);
    \draw[thick] (-4.798, 2.54) -- (-4.798, 1.129);
    \draw[RedOrange, thick] (-6.491, 1.976) -- (-4.233, 1.976);
    \draw[RedOrange, thick] (-6.491, 1.411) -- (-4.233, 1.411);
    \node[disk normal] at (-5.927, 2.258) {};
    \node[disk normal, Green!80] at (-5.362, 1.976) {};
    \node[disk normal] at (-4.798, 2.258) {};
    \node[disk normal, Green!80] at (-5.362, 1.411) {};
    \node[disk normal] at (-5.927, 1.693) {};
    \node[disk normal] at (-4.798, 1.693) {};
    \node[anchor=center] at (-6.221, 2.281) {$L_g$};
    \node[anchor=center] at (-4.498, 2.312) {$L_g$};
    \node[anchor=center] at (-6.246, 1.696) {$L_h$};
    \node[anchor=center] at (-4.489, 1.709) {$L_h$};
    \node[anchor=center] at (-5.34, 2.227) {$\omega_g$};
    \node[anchor=center] at (-5.372, 1.169) {$\omega_h$};
    \node[anchor=center] at (-3.953, 1.733) {$=$};
    \draw[RedOrange, thick] (-3.387, 1.976) -- (-2.822, 1.693) -- (-3.387, 1.411);
    \draw[RedOrange, thick] (-2.822, 1.693) -- (-2.258, 1.693);
    \draw[thick] (-2.54, 2.258) -- (-2.54, 1.129);
    \draw[RedOrange, thick] (-2.258, 1.693) -- (-1.693, 1.976);
    \draw[RedOrange, thick] (-2.258, 1.693) -- (-1.693, 1.411) -- (-1.129, 1.411) -- (-0.564, 1.693) -- (-1.129, 1.976) -- (-1.693, 1.976);
    \draw[thick] (-0.282, 2.258) -- (-0.282, 1.129);
    \draw[RedOrange, thick] (0.564, 1.976) -- (0, 1.693) -- (0.564, 1.411);
    \node[disk normal, Green!80] at (-1.411, 1.976) {};
    \node[disk normal, Green!80] at (-1.411, 1.411) {};
    \node[disk normal, Green!80] at (-1.411, 1.976) {};
    \node[disk normal] at (-2.54, 1.976) {};
    \node[disk normal] at (-0.282, 1.976) {};
    \node[disk normal] at (-2.258, 1.693) {};
    \node[disk normal] at (-2.822, 1.693) {};
    \node[disk normal] at (-0.564, 1.693) {};
    \node[disk normal] at (0, 1.693) {};
    \node[anchor=center] at (-2.832, 2.181) {$L_{gh}$};
    \node[anchor=center] at (0.091, 2.172) {$L_{gh}$};
    \node[anchor=center] at (-1.403, 2.2) {$\omega_g$};
    \node[anchor=center] at (-1.389, 1.172) {$\omega_h$};
    \node[disk normal] at (-3.224, 1.9) {};
    \node[disk normal] at (-1.9, 1.897) {};
    \node[disk normal] at (-0.932, 1.887) {};
    \node[disk normal] at (0.359, 1.879) {};
    \node[anchor=center, font=\scriptsize] at (-1.921, 0.794) {$\omega(g,h,r)$};
    \node[anchor=center, font=\scriptsize] at (-3.287, 0.826) {$\bar{\omega}(g,h,r)$};
    \node[anchor=center, font=\scriptsize] at (-0.456, 0.83) {$\bar{\omega}(g,h,s)$};
    \node[anchor=center, font=\scriptsize] at (0.749, 0.843) {$\omega(g,h,s)$};
    \node[anchor=center, font=\scriptsize] at (-3.466, 1.75) {$L_h$};
    \node[anchor=center, font=\scriptsize] at (-1.927, 2.121) {$L_{\bar{h}}$};
    \node[anchor=center, font=\scriptsize] at (-0.862, 2.087) {$L_h$};
    \node[anchor=center, font=\scriptsize] at (0.604, 1.724) {$L_{\bar{h}}$};
    \draw[black!50, thick, -Latex] (-2.849, 1.623) -- (-3.369, 1.071);
    \draw[black!50, thick, -Latex] (-2.227, 1.592) -- (-2.193, 1.049);
    \draw[black!50, thick, -Latex] (-0.574, 1.59) -- (-0.586, 1.123);
    \draw[shift={(0.042, 1.584)}, xscale=1.418, yscale=1.269, black!50, thick, -Latex] (0, 0) -- (0.271, -0.405);
    \node[anchor=center, font=\footnotesize] at (-2.547, 1.031) {$r$};
    \node[anchor=center, font=\footnotesize] at (-0.276, 1.053) {$s$};
    \node[anchor=center, font=\footnotesize] at (-4.798, 1.031) {$s$};
    \node[anchor=center, font=\footnotesize] at (-5.911, 1.026) {$r$};
    \draw[black!50, thick] (-2.093, 2.294) .. controls (-1.652, 2.5) and (-1.317, 2.503) .. (-0.726, 2.307);
    \node[anchor=center, font=\scriptsize] at (-1.427, 2.64) {${\omega}(g,hr,r^{-1}s)$};
    \node[anchor=center] at (1.331, 1.755) {$=$};
    \draw[RedOrange, thick] (2.147, 2.063) -- (2.711, 1.781) -- (2.147, 1.499);
    \draw[RedOrange, thick] (2.711, 1.781) -- (3.276, 1.781);
    \draw[thick] (2.994, 2.346) -- (2.994, 1.217);
    \node[disk normal] at (2.994, 2.063) {};
    \node[disk normal, Green!80] at (3.276, 1.781) {};
    \node[disk normal] at (2.711, 1.781) {};
    \node[anchor=center] at (2.701, 2.269) {$L_{gh}$};
    \node[disk normal] at (2.31, 1.988) {};
    \node[anchor=center, font=\scriptsize] at (2.455, 0.878) {$\bar{\omega}(g,h,r)$};
    \node[anchor=center, font=\scriptsize] at (2.068, 1.837) {$L_h$};
    \draw[shift={(2.685, 1.711)}, xscale=0.535, yscale=1.086, black!50, thick, -Latex] (0, 0) -- (-0.52, -0.553);
    \node[anchor=center, font=\footnotesize] at (2.987, 1.119) {$r$};
    \draw[RedOrange, thick] (3.835, 1.795) -- (4.399, 1.513);
    \draw[thick] (3.558, 2.346) -- (3.558, 1.217);
    \node[disk normal] at (3.558, 2.063) {};
    \node[disk normal] at (3.84, 1.781) {};
    \node[anchor=center, font=\footnotesize] at (3.551, 1.119) {$s$};
    \node[anchor=center] at (3.908, 2.236) {$L_{gh}$};
    \node[disk normal] at (4.166, 1.96) {};
    \node[anchor=center, font=\scriptsize] at (4.432, 1.824) {$L_{\bar{h}}$};
    \node[anchor=center] at (3.293, 1.573) {$\omega_{gh}$};
    \node[anchor=center, font=\scriptsize] at (4.564, 0.889) {$\omega(g,h,s)$};
    \draw[shift={(3.866, 1.696)}, xscale=1.418, yscale=1.269, black!50, thick, -Latex] (0, 0) -- (0.271, -0.405);
\end{tikzpicture}\ ,
\end{equation}
where we have used the 3-cocycle equation to compress the middle bubble and we have simplified $h^{-1}$ by $\bar{h}$.
{\bf Unitarity:} So far we have shown that actually $U_g U_h = U_{gh}$. Using the normalized version of a $3$-cocycle where $\omega=1$ if any of its entries is the trivial group element $1$, it is easy to see that $U_1 = \id$ and then $U^{-1}_g= U_{g^{-1}}$. We now proceed to show that $U^\dagger_g = U_{g^{-1}}$. We first write down the site tensor of the operator $U^\dagger_g$ which reads:
\begin{equation}\label{Tdagger}
\begin{tikzpicture}[scale=1]
    \draw[shift={(-3.885, 2.264)}, xscale=1.39, yscale=3.456, RedOrange, thick] (0, 0) -- (1.411, 0);
    \draw[thick] (-3.057, 2.828) -- (-3.057, 1.699);
    \node[disk normal] at (-2.797, 2.262) {};
    \node[disk normal, Green!80] at (-2.276, 2.264) {};
    \node[anchor=center] at (-3.563, 1.963) {$L_{\bar{g}}$};
    \node[anchor=center] at (-2.27, 2.523) {$\bar{\omega}_g$};
    \node[anchor=center] at (-2.721, 1.966) {$L_g$};
    \node[disk normal] at (-3.064, 2.602) {};
    \node[disk normal] at (-3.466, 2.264) {};
    \draw[RedOrange, thick] (-5.96, 2.251) -- (-4.549, 2.251);
    \draw[thick] (-5.395, 2.815) -- (-5.395, 1.687);
    \node[disk normal] at (-5.402, 1.957) {};
    \node[disk normal, Green!80] at (-4.831, 2.251) {};
    \node[anchor=center] at (-5.694, 1.836) {$L_{\bar{g}}$};
    \node[anchor=center] at (-4.89, 2.538) {$\bar{\omega}_g$};
    \node[anchor=center] at (-3.363, 2.637) {$L_{\bar{g}}$};
    \node[anchor=center] at (-4.213, 2.261) {$=$};
    \draw[shift={(-1.31, 2.261)}, xscale=1.39, yscale=3.456, RedOrange, thick] (0, 0) -- (1.411, 0);
    \draw[thick] (-0.482, 2.825) -- (-0.482, 1.696);
    \node[disk normal] at (0.366, 2.255) {};
    \node[disk normal, Green!80] at (-0.171, 2.26) {};
    \node[anchor=center] at (-0.988, 1.96) {$L_{\bar{g}}$};
    \node[anchor=center] at (-0.125, 1.995) {$\hat{\omega}_g$};
    \node[anchor=center] at (0.419, 1.965) {$L_g$};
    \node[disk normal] at (-0.489, 2.599) {};
    \node[disk normal] at (-0.891, 2.261) {};
    \node[anchor=center] at (-0.788, 2.635) {$L_{\bar{g}}$};
    \node[anchor=center] at (-1.638, 2.258) {$=$};
\end{tikzpicture} \ ,
\end{equation}
where the new introduced matrix is $(\hat{\omega}_g)_{k,l} = \bar{\omega}(g,g^{-1}k, k^{-1}l)$. Using the $3$-cocycle equation \eqref{3cocyleeq} with $(g,h,k,l) = (g,g^{-1},k,k^{-1}l)$ we can see that:
$$ \bar{\omega}(g,g^{-1}k, k^{-1}l) = {\omega}(g^{-1},k, k^{-1}l) \omega(g,g^{-1},k) \bar{\omega}(g,g^{-1},l) \ . $$
Then, by defining the diagonal matrix $(D_g)_{k,l}= \delta_{k,l} \omega(g,g^{-1},k) $ with $D_g \bar{D}_g= \id$ we can write 
$$\hat{\omega}_g = D_g \cdot \omega_{g^{-1}}\cdot \bar{D}_g \ , $$
which implies that the tensor of $U^\dagger_g$  in Eq.\eqref{Tdagger} is related by a local gauge with $T_{g^{-1}}$ which implies that $U^\dagger_g = U_{g^{-1}}$.\\

The $(2+1)$ SPT state is best seen by placing it on a chessboard pattern with $\mathbb{C}[G]$ Hilbert spaces in the edges. The sites corresponds to the four edges of the square of one of the sublattices and the plaquettes to the other sublattice. The state is simply a generalized 4-party $GHZ$-state, $\sum_g\ket{ gggg}$, on each plaquette:
$$

$$

The on-site unitary of the global symmetry of that state is the following one:
$$
%
$$
which satisfies the group relations by virtue of Eq.~\eqref{tgthdecomp}. We note that the change of $\omega$ to $\bar{\omega}$ on half of the virtual level of $u_g$ is necessary to show that it is actually a global symmetry of the state. Concretely, it allows us to push the action of the on-site operators to the virtual boundary of the applied region. To see this, we first compute what is the MPO that goes to the virtual level when applied to a single site of the state:

$$
%
 \  \ ,
$$
where the orientation of the tensor network dictates the application of $L_g$ or $L_{\bar{g}}$. As depicted, the shape of the virtual MPO symmetry is a truncated square. The MPO tensor is:
$$
T'_g = 
%
\ ,
$$
which implies that the MPO constructed with it, $O'_g$,  is not unitary: $O'_gO'_{\bar{g}}= O'_e$ is a projector onto the tensor product of maximally entangled pair states. Finally, this virtual MPO grows correctly when applied to neighboring sites due to the following calculation:
$$
%
 \ ,
$$
so the boundary MPO carries a $3$-cocycle index.\\

\subsection{(3+1) SPT phases based on TN}

The three-dimensional lattice where we place our state is a cubic-lattice with the faces valued on the Hilbert space $\mathbb{C}[G]$. We split the cubes in two sublattices where a sublattice corresponds to the sites and the other sublattice corresponds to the plaquettes. We think of the six faces of each site or plaquette as the vertices of an octahedron:

\begin{equation}\label{3dlat}
%
 
 \ , 
\end{equation}
as such the lattice is a tetrahedral-octahedral honeycomb.

The 3D state we consider is  the tensor product of generalized 6-party $GHZ$-states, $\sum_g\ket{ gggggg}$, on each plaquette:

$$
\ket{\Psi_{3D}} = \bigotimes_{p} \left (
%
\right )_p
$$

The on-site unitary operator that forms the global 3D symmetry is based on the 2D unitary of Sec.~\ref{secTNU}: The on-site TN operator acts on the vertices of the octahedron formed by the cube's faces and the phase factors $\omega$ and $\bar{\omega}$ are associated to the faces of the octahedron:
\begin{equation}\label{u_g}
u_g =
%
\end{equation}

To see that this on-site unitary indeed defines a global symmetry of the state $\ket{\Psi_{3D}} $, we will show that the corresponding virtual TNO grows correctly when concatenated with the surrounding sites. The first thing to notice is that each vertex of the octahedron, corresponding to a site, is connected with an octahedron corresponding to a plaquette:
$$
%
 \ ,
$$
and that each vertex of the octahedra is connected to four of the other vertices. This shows that the physical unitary $u_g$ with the shape of an octahedron is pushed to the virtual level as a TNO with the shape of a truncated octahedron:
$$
o_g = 
%
 
\ ,
$$
where the action of a single $L_g$ in $u_g$ is translated into four matrices, $L_g$ or $L^\dagger_g$ depending on the orientation of the square faces in $o_g$. This orientation is chosen to allow for an appropriate growing of the concatenation of neighboring virtual symmetry operators $o_g$.

\section{Conclusions and Outlook}

In this work we have proposed local tensor equations that are sufficient to construct 2D representations of finite groups with an anomalous index, a $4$-cocycle. We have used them to study (2+1)D gapped  phases invariant under those anomalous symmetries using PEPS as its ground states. We have also studied (3+1) SPT states invariant under on-site global symmetries whose boundary symmetry defects are realized by our proposed TN operators.

It is worth to mention that general symmetries in 2D are given by fusion 2-categories \cite{Inamura_2024}. Under that viewpoint, in this paper we have studied $\mathsf{2Vec}_G^\omega$ symmetries, where $\omega$ is the $4$-cocycle corresponding to the 10-j symbol and the solutions of the mixed cocycle equation \eqref{mixed4cocycle} are associated with the module 2-categories over the fusion 2-category of the symmetry where our $\Lambda$ is the module 10-j symbols \cite{inamura202521dlatticemodelstensor}. A natural generalization of our work would be to extend our local tensor equations to fusion 2-categories.

Given a TN representation of a group we can use it to construct topologically ordered states in one dimension higher in the spirit of Ref.~\cite{Schuch10}. As such, we propose the following site tensor for the ground state of a $(3+1)D$ topologically ordered state  based on $G$ twisted by a $4$-cocycle.
$$
\begin{tikzpicture}[scale=0.75]
    \draw[RoyalBlue!80, very thick] (-2.901, 2.158) -- (-2.104, 2.377);
    \draw[RoyalBlue!80, very thick] (-1.693, 1.411) -- (-1.129, 0.847);
    \draw[RoyalBlue!80, very thick] (-0.564, 2.54) -- (0, 1.976);
    \draw[very thick, thick bevel] (-3.669, 3.387) rectangle (-1.411, 1.129);
    \draw[very thick] (-2.54, 4.516) rectangle (-0.282, 2.258);
    \draw[very thick] (-3.669, 3.387) -- (-2.54, 4.516);
    \draw[very thick] (-1.411, 1.129) -- (-0.282, 2.258);
    \draw[very thick] (-1.411, 3.387) -- (-0.282, 4.516);
    \draw[very thick] (-2.54, 2.258) -- (-3.669, 1.129);
    \draw[RoyalBlue!80, very thick] (-3.951, 3.669) -- (-3.387, 3.104);
    \draw[RoyalBlue!80, very thick] (-2.822, 4.798) -- (-2.258, 4.233);
    \draw[RoyalBlue!80, very thick] (-0.697, 4.311) -- (0.115, 4.701);
    \draw[RoyalBlue!80, very thick] (-1.759, 3.112) -- (-1.002, 3.649);
    \draw[RoyalBlue!80, very thick] (-3.969, 0.956) -- (-3.182, 1.409);
    \node[disk normal] at (-3.66, 3.378) {};
    \node[disk normal] at (-2.523, 4.498) {};
    \node[disk normal] at (-0.282, 4.483) {};
    \node[disk normal] at (-1.41, 3.388) {};
    \node[disk normal] at (-1.405, 1.135) {};
    \node[disk normal] at (-0.275, 2.251) {};
    \node[disk normal] at (-3.654, 1.144) {};
    \node[disk normal] at (-2.523, 2.258) {};
    \node[anchor=center, font=\Large] at (-5, 2.659) {$\frac{1}{|G|}\sum_g $};
    \node[anchor=center] at (-1.575, 3.7) {$T_g$};
\end{tikzpicture} \ ,
$$
where the tensor $T_g$ is the one used in \eqref{Og}. This tensor has naturally a virtual global $G$-symmetry. This kind of TN representation has been studied before in the toric code model in Ref.~\cite{Delcamp21} and generalize for any group in Ref.\cite{Delcamp_2022} for trivial $4$-cocycle. We wonder if the 'pushing through' equations are also satisfied in these models and if a generalization to fusion $2$-categories is possible. Moreover, it would be very interesting to see if the other TN representation described there, with loop symmetries instead of global ones \cite{Williamson21B}, can be generalized when considering a non-trivial $4$-cocycle and in general a fusion 2-category.

Some questions remain open. For example, we wonder if the 'inner product' of four PEPS as defined in \cite{Shuhei2D} could be used here to obtain the invariant of Eq.~\eqref{4simplexLambda}, similar to what has been done for the one-dimensional case \cite{inamura202411dsptphasesfusion}. Another interesting point would be to study the symmetry defects of these anomalous symmetries in 2D since already interesting phenomena in 1D appeared \cite{Garre24_DWMPU}.

\section*{Acknowledgments}
JGR thanks Clement Delcamp for the very useful explanations on his work.
JGR is funded by the FWF Erwin
Schr\"odinger Program (Grant DOI 10.55776/J4796).
AM is funded in part by the European Union’s Horizon 2020 research and innovation program through Grant No. 863476 (ERC-CoG SEQUAM).

\bibliography{bibliography}
\end{document}